\documentclass[11pt]{article}
\usepackage{graphicx}
\usepackage{amssymb}
\usepackage{graphics}
\usepackage{array}

\input{tcilatex}
\begin{document}

\title{Cooling neutrons using non-dispersive magnetic excitations }
\author{O. Zimmer \\
Institut Laue Langevin, 38042 Grenoble, France}
\maketitle

\begin{abstract}
A new method is proposed for cooling neutrons by inelastic magnetic
scattering in weakly absorbing, cold paramagnetic systems. Kinetic neutron
energy is removed in constant decrements determined by the Zeeman energy of
paramagnetic atoms or ions in an external magnetic field, or by zero-field
level splittings in magnetic molecules. Analytical solutions of the
stationary neutron transport equation are given using inelastic neutron
scattering cross sections derived in an appendix. They neglect any inelastic
process except the paramagnetic scattering and hence still underestimate
very-cold neutron densities. Molecular oxygen with its triplet ground state
appears particularly promising, notably as a host in fully deuterated O$_{2}$%
-clathrate hydrate, or more exotically, in dry O$_{2}$-$^{4}$He van der
Waals clusters. At a neutron temperature about $6\ \mathrm{K}$, for which
neutron conversion to ultra-cold neutrons by single-phonon emission in pure
superfluid $^{4}$He works best, conversion rates due to paramagnetic
scattering in the clathrate are found to be a factor $9$ larger. While in
conversion the neutron imparts only a single energy quantum to the medium,
the multi-step paramagnetic cooling cascade leads to further strong
enhancements of very-cold neutron densities, e.g., by a factor $14$ ($57$)
for an initial neutron temperature of $30\ \mathrm{K}$ ($100\ \mathrm{K}$),
for the moderator held at about $1.3\ \mathrm{K}$. Due to a favorable Bragg
cutoff of the O$_{2}$-clathrate the cascade-cooling can take effect in a
moderator with linear extensions smaller than a meter. The paramagnetic
cooling mechanism may offer benefits in novel intense sources of very cold
neutrons and for enhancing production of ultra-cold neutrons.\bigskip

Keywords: neutron scattering, neutron moderation, very cold neutrons,
ultra-cold neutrons
\end{abstract}

\section{Introduction}

Thermal and cold neutrons play an important role for fundamental research,
on one hand as a probe to study condensed matter systems and for inducing
nuclear reactions, on the other hand as an object and a tool for studying
low-energy particle physics via precise measurements of the static and decay
properties of the neutron. Cold neutrons are commonly produced by slowdown
from thermal energies, using a cold medium with a suitable dynamical
structure factor and weak absorption. Liquid hydrogen and deuterium held at
the respective boiling points at about $20$ and $24$ K are particularly
effective for moderation and have gained practical importance as "cold
neutron sources" implemented in high-flux neutron research facilities. For
instance, a liquid-deuterium cold source at the Institut Laue Langevin (ILL)
in Grenoble, France, provides a gain factor $80-100$ for low-energy neutrons
with wavelengths $\lambda >1\ \mathrm{nm}$ \cite{Ageron/1989}. For further
cooling of neutrons one would need a colder medium with suitable degrees of
freedom. Collective excitations such as phonons in a cold solid with Debye
temperature $T_{\mathrm{D}}$ are however rather inappropriate for moderation
since for neutron energy, $E<k_{\mathrm{B}}T_{\mathrm{D}}$, the cross
section for single phonon emission is proportional to $\left( E/k_{\mathrm{B}%
}T_{\mathrm{D}}\right) ^{3}$, and cross sections for higher order processes
drop even faster with decreasing energy \cite{Gurevich/1968}. Localized,
dispersion-free excitations on the other hand do not suffer from this
limitation and may work even at lowest neutron energies.

A peculiar, forty years old proposal of "phononless cooling of neutrons to
extremely low temperatures" involves tiny energy transfers in units of
neutron and nuclear Zeeman energy \cite{Namiot/1974}. In a moderator
comprized of deuterons with nuclear spins highly polarized along a strong
magnetic field, conservation of angular momentum allows simultaneous neutron
and nuclear spin flips to happen only in the doublet state and hence, due to
the associated positive change of Zeeman energies, only with the removal of
neutron kinetic energy. Since even in magnetic fields of several Tesla this
amounts to less than a $\mathrm{\mu eV}$, many spin flip collisions will be
necessary for significant neutron cooling. The author of the proposal
pointed out that, even for an ideally polarized deuteron system in a field
as strong as $30$ T, cooling to extremely low temperatures would take effect
only after preliminary cooling of the neutron spectrum to at least $12$ K.
These conditions explain why the proposal has not yet been put into
practice. A similar situation appears in refrigeration of bulk matter to
lowest temperatures using adiabatic nuclear demagnetization. A popular
method to achieve the necessary precooling is adiabatic demagnetization of
paramagnetic salts, which due to the large electronic magnetic moments is
much better adapted for cooling at a higher temperature than a nuclear stage.

Guided by this analogy we are led to consider slowdown of neutrons due to
scattering by a paramagnetic system at low temperature, which has not yet
been reported in the literature. Here, the transfer of energy in a
scattering process with electron spin flip is typically three orders of
magnitude larger than for nuclear spin flip scattering, due to the ratio of
the Bohr and nuclear magnetons. For simple paramagnetic atomic or ionic
species it is determined by an external field, whereas magnetic molecules
possess zero-field splittings of magnetic energy levels able to remove
neutron kinetic energy even without an external field. As a consequence of
the non-dispersivity of paramagnetic excitations, neutron cooling can
proceed in inelastic, incoherent scattering cascades with energy decrement $%
E^{\ast }$ of typically a fraction of a \textrm{meV}.

A particular motivation for this study is the quest for efficient production
methods of neutrons in the low-energy range of the spectrum provided by a
cold source, commonly called very cold neutrons (VCN) and ultra-cold
neutrons (UCN). More intense beams of VCN would enhance capabilities of
neutron scattering techniques such as reflectometry, spin echo spectroscopy
and interferometry, to mention only some classical applications. They may
also offer new opportunities for fundamental physics projects, such as
beam-based searches for a non-vanishing neutron electric dipole moment (EDM) 
\cite{Piegsa/2013}, for neutron-antineutron oscillations \cite{Babu/2013},
or for new fundamental forces \cite{Piegsa/2012}. UCN on the other hand can
be trapped in bottles made of materials with positive neutron optical
potential, by magnetic field gradients, and by gravity \cite%
{Golub/1991,Ignatovich/1990}. Owing to this peculiar property they have
become a valuable tool for a plethora of investigations in fundamental
physics \cite{Dubbers/2011,Musolf/2008,Abele/2008}.

A classical method for production of UCN and VCN at the ILL extracts
low-energy neutrons from a liquid-deuterium moderator through a vertical
neutron guide, followed by a phase space transformation to lowest energies
using a neutron turbine \cite{Steyerl/1986}. More recently, alternative
methods have begun to provide competitive UCN densities \cite%
{Lauer/2013,Frei/2007,Saunders/2013,Masuda/2012,Lauss/2012,Zimmer/2011,Zimmer/2010}%
. They involve conversion of cold neutrons in single inelastic scattering
events \cite{Golub/1975}, instead of multiple energy transfers
characteristic for neutron slowdown in a moderator. In its simplest form the
converter acts as an effective two-level system with an excited state
separated by an energy $E^{\ast }$ from the ground state. At temperatures $%
k_{\mathrm{B}}T\ll E^{\ast }$ the excited state is strongly depleted,
leading to a suppression of neutron up-scattering back to higher energy.
Converter materials investigated so far are superfluid helium \cite%
{Golub/1977}, solid deuterium \cite{Golub/1977,Golub/1983}, solid $\alpha $%
-oxygen \cite{Gutsmiedl/2011,Liu/2004} and solid $^{15}$N \cite{Salvat/2013}%
. All cases employ phonons (with a contribution of magnons for $\alpha $%
-oxygen), with $E^{\ast }$ in the order of one to several meV (e.g., $1$ meV
for superfluid $^{4}$He).

For paramagnetic scattering typical single energy transfers are smaller
(e.g., $0.4$ meV for molecular oxygen encaged in the inclusion compounds
discussed below). More importantly and contrary to dispersive, collective
excitations, energy transfers can be cascaded over a comb of many
equidistant neutron energy groups within a broad incident neutron spectrum.
Here we show that this paramagnetic cooling cascade provides an efficient
channel for low-energy neutron moderation prior to a final conversion
process to UCN or VCN, leading to a large enhancement of conversion rates.
Particularly promising are materials involving molecular oxygen. Besides
other advantages they can be kept close to the magnetic ground state at
ordinary liquid-helium temperatures, which is a helpful feature for
technical implementations.

\section{Neutron conversion by a paramagnetic electron spin system}

We start with a discussion of a paramagnetic electron spin system as a
neutron converter, i.e.\ neglecting multiple inelastic scattering events.
This situation prevails if only a small amount of material is exposed to a
neutron field. Key criteria are a large density of unpaired electrons and
weak neutron absorption. Paramagnetic atomic and molecular species
worthwhile to be considered are listed in Table $\ref{Species}$. They
involve the nuclides $^{2}$H, $^{16}$O and $^{15}$N that possess absorption
cross sections $\sigma _{\mathrm{a}}<10^{-27}\ \mathrm{cm}^{2}$ for thermal
neutrons and can be prepared in systems with paramagnetic spin densities
exceeding $10^{20}\ \mathrm{cm}^{-3}$. Pure $^{4}$He is the only existing
medium with no absorption at all. It possesses paramagnetic states in form
of single-electron bubbles and the He$_{2}^{\ast }\left( a^{3}\Sigma _{%
\mathrm{u}}^{+}\right) $ excimer triplet state. Sufficient bulk density is
however an obvious issue for charged or unstable species. On the other hand,
the vanishing absorption makes $^{4}$He an interesting matrix for hosting
paramagnetic atoms and molecules. 
\begin{table}[tbp] \centering%
\begin{tabular}{|l|l|l|l|}
\hline
Species & $S$ & $g_{-}\left( T\rightarrow 0\right) $ & $\sigma _{\mathrm{a}}%
\text{ (}\mathrm{mbarn}\text{)}$ \\ \hline
$\text{electron}$ & $1/2$ & $1/3$ & $0$ \\ 
$^{2}\text{H}$ & $1/2$ & $1/3$ & $0.519\left( 7\right) $ \\ 
$^{1}\text{H}$ & $1/2$ & $1/3$ & $332.6\left( 7\right) $ \\ 
$^{15}\text{N}$ & $3/2$ & $1$ & $0.024\left( 8\right) $ \\ 
$^{14}\text{N}$ & $3/2$ & $1$ & $1910\left( 30\right) $ \\ 
$^{16}\text{O}_{2}$ & $1$ & $4/3$ & $2\times 0.10\left( 2\right) $ \\ 
$\text{natural O}_{2}$ & $1$ & $4/3$ & $2\times 0.19\left( 2\right) $ \\ 
\hline
\end{tabular}%
\caption{Weakly absorbing paramagnetic species with electronic spin $S$. For atomic hydrogen and nitrogen
also the strongly absorbing isotopic contaminants are quoted. The cross section for magnetic down-scattering
is proportional to the thermal factor $g_{-}\left( T\right) $ (see eq.\ \ref{Sigma+---+} and the appendix). Neutron 
absorption cross sections are quoted for neutrons with a speed of $2200$ m/s.}%
\label{Species}%
\end{table}%

For calculation of conversion rates we need cross sections for neutron
scattering with an electron spin flip without neglecting the change of
neutron kinetic energy due to Zeeman or molecular zero-field splittings. As
they seem not to appear in the literature\footnote{%
Magnetic neutron scattering theory serves for analyzing experimental data
for investigating structure and dynamics of condensed matter systems. In
paramagnetic systems the scattering associated with an electron spin flip is
diffuse and thus of rather limited interest. Even in presence of magnetic
fields it is usually treated as elastic, neglecting small changes of neutron
kinetic energy due to Zeeman splittings \cite{Lovesey/1984,Squires/1978}.}
we derive them in the appendix. We write the macroscopic,
energy-differential cross section of a paramagnetic system for neutron
scattering from an initial energy $E$ (wavenumber $k$) to a final energy $%
E^{\prime }$ (wavenumber $k^{\prime }$) as 
\begin{equation}
\Sigma ^{\pm }\left( E\rightarrow E^{\prime }\right) =n_{\mathrm{pc}}\sigma
_{\mathrm{m}}\frac{k^{\prime }}{k}g_{\pm }\left( T\right) f_{\pm }\left(
E\right) \delta \left( E\pm E^{\ast }-E^{\prime }\right) ,
\label{Sigma+---+}
\end{equation}%
where the upper (lower) sign stands for a process with neutron energy gain
(loss), $n_{\mathrm{pc}}$ is the number density of paramagnetic centers of a
single species with spin $S$, and $\sigma _{\mathrm{m}}=4\pi b_{\mathrm{m}%
}^{2}\approx 3.66\ \mathrm{barn}$ with the magnetic scattering length as
defined in eq.\ \ref{b_m} in the appendix. For the dispersion-free
excitations considered here the (positive) transfer energy $E^{\ast }$ is
independent on the neutron energy and balances a corresponding loss or gain
in neutron kinetic energy. Obviously, neutron down-scattering is only
possible for $E>E^{\ast }$. For paramagnetic atoms or ions with g-factor $g$
and without crystal field splittings, 
\begin{equation}
E^{\ast }=\left\vert g\mu _{\mathrm{B}}B_{0}\right\vert =\frac{\left\vert
g\right\vert }{2}115.8\ \mathrm{\mu eV}\times B_{0}\left[ \mathrm{T}\right]
\label{E*-Zeeman}
\end{equation}%
is the Zeeman splitting of magnetic states in an external magnetic field $%
B_{0}$. For the species quoted in Table $\ref{Species}$, $g\approx -2$. With
respect to the electronic Zeeman energy the neutron Zeeman energy is
negligible and not taken into account here. Molecular oxygen has a
paramagnetic spin triplet ground state ($^{3}\Sigma _{\mathrm{g}}^{-}$).
Without external magnetic field, 
\begin{equation}
E^{\ast }=D,  \label{E*-oxygen}
\end{equation}%
where $D$ is the zero-field splitting constant of the spin $S=1$ states with
projection $m=0$ and $m=\pm 1$ along the molecular axis. Using electron spin
resonance spectroscopy $D=491\ \mathrm{\mu eV}$ was measured for oxygen in
the gas phase \cite{Tinkham/1955}. Electrostatic effects on O$_{2}$
molecules embedded in a bulk matter matrix may modify this value, as
noticeable in the systems discussed below. The functions $g_{\pm }\left(
T\right) $ contain thermal averages of spin matrix elements and are given in
the appendix. In the low temperature limit, $k_{\mathrm{B}}T\ll E^{\ast }$,
where the system is close to its magnetic ground state,%
\begin{equation}
g_{+}\left( T\rightarrow 0\right) \rightarrow 0
\end{equation}%
leads to suppression of neutron up-scattering, while cross sections for
down-scattering become proportional to%
\begin{equation}
g_{-}\left( T\rightarrow 0\right) \rightarrow \left\{ 
\begin{array}{ll}
2S/3\quad & \left( \text{Zeeman system}\right) \\ 
4/3 & \left( \text{O}_{2}\text{ molecule}\right)%
\end{array}%
\right. .
\end{equation}%
The functions $f_{\pm }\left( E\right) $ in eq.\ \ref{Sigma+---+} account
for the effect of the magnetic form factor $F\left( \mathbf{\kappa }\right) $%
, which depends on the transfer wave vector $\mathbf{\kappa =k-k}^{\prime }$%
. A theoretical expression of the magnetic form factor of gaseous oxygen
based on Meckler's electron wave functions \cite{Meckler/1953} was given by
Kleiner \cite{Kleiner/1955} and found to be very similar to the form factor
measured in condensed phases \cite{Deraman/1985, Meier/1984}. For encaged,
unoriented O$_{2}$ molecules we may write, following a discussion in ref.\ 
\cite{Kleiner/1955},%
\begin{equation}
f_{\pm }\left( E\right) =\frac{1}{2}\int_{0}^{\pi }\left\langle \left\vert
F\right\vert ^{2}\right\rangle \left( \kappa _{\pm }\left( E,\theta \right)
\right) \sin \theta d\theta ,  \label{form factor effect}
\end{equation}%
where the brackets stand for orientational averaging of the molecules. For
our purposes we approximate $\left\langle \left\vert F\right\vert
^{2}\right\rangle $ by a Gaussian with a half width of $15\ \mathrm{nm}^{-1}$
at half maximum. The functional dependence of $\kappa _{\pm }$ on the
initial neutron energy $E$ and the scattering angle $\theta $ is given by%
\begin{equation}
\frac{\hbar ^{2}}{2m_{\mathrm{n}}}\kappa _{\pm }^{2}=2E\pm E^{\ast }-2\sqrt{%
E\left( E\pm E^{\ast }\right) }\cos \theta .
\end{equation}%
Figure $1$ shows $f_{-}\left( E\right) $. Note that $f_{+}\left( E\right)
=f_{-}\left( E+E^{\ast }\right) $. Equation\ \ref{Sigma+---+} should also
contain the Debye-Waller factor, which for a harmonically bound center with
mass $M$ and oscillation frequency $\omega _{0}$ can be written as (see,
e.g., ref.\ \cite{Gurevich/1968})%
\begin{equation}
\exp \left( -2W\right) =\exp \left( -\frac{\hbar \kappa ^{2}}{2M\omega _{0}}%
\right) .
\end{equation}%
However, for $M\gg m_{\mathrm{n}}$ and a cold system where the oscillators
are in the ground state, it reduces scattering cross sections by less than a
few per cent and is therefore neglected in the further analysis.

\begin{figure}[tbp]
\centering
\includegraphics[width=0.8\textwidth]{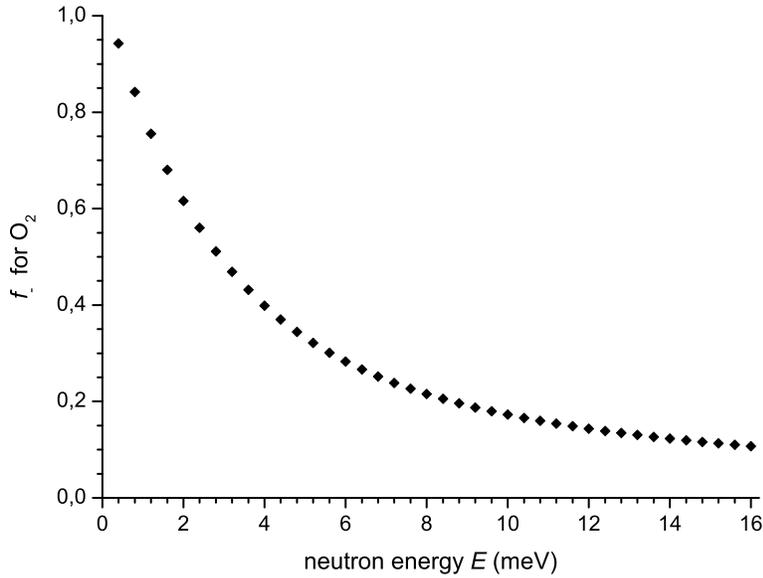}
\caption{The function $f_{-}\left( E\right) $ as
defined in eq.\ \protect\ref{form factor effect}, describing the effect of
the magnetic form factor of molecular oxygen on the neutron scattering cross
section.}
\label{fig:figure1}
\end{figure}

The spectral conversion rate density, i.e.\ the number of neutrons converted
in a neutron field to energy $E^{\prime }<E$, per units of time, volume and
energy, is given by%
\begin{equation}
p\left( E^{\prime }\right) =\int_{0}^{\infty }\Phi \left( E\right) \Sigma
^{-}\left( E\rightarrow E^{\prime }\right) dE,  \label{UCN prod rate density}
\end{equation}%
where $\Phi \left( E\right) dE$ denotes a spatially homogeneous incident
flux of neutrons with energies in an interval of width $dE$ about $E$ \cite%
{Beckurts/1964}. Using eq.\ \ref{Sigma+---+} and the relation%
\begin{equation}
\Phi \left( E\right) =n\left( E\right) v\left( E\right) ,
\end{equation}%
with the spectral neutron density $n\left( E\right) $ and the neutron speed $%
v\left( E\right) =\sqrt{2E/m_{\mathrm{n}}}$ ($m_{\mathrm{n}}$ is the neutron
mass), one obtains%
\begin{equation}
p\left( E^{\prime }\right) =n\left( E^{\prime }+E^{\ast }\right) n_{\mathrm{%
pc}}\sigma _{\mathrm{m}}g_{-}\left( T\right) f_{-}\left( E^{\prime }+E^{\ast
}\right) v\left( E^{\prime }\right) .  \label{p(E')}
\end{equation}%
For the special case of conversion to UCN, we note that their energy range, $%
0<E^{\prime }\leq E_{\mathrm{c}}$, is defined by a cutoff $E_{\mathrm{c}}$,
normally with a value less than $350\ \mathrm{neV}$, e.g.\ due to the
neutron optical potential of the wall material of a UCN bottle. Comparison
with eq.\ \ref{E*-oxygen}, or with eq.\ \ref{E*-Zeeman} for not too small
magnetic field, shows that $E^{\ast }\gg E_{\mathrm{c}}$. The total rate
density of neutron conversion to UCN follows simply from integration of eq.\ %
\ref{p(E')} over all UCN energies, i.e. 
\begin{equation}
p=\int_{0}^{E_{\mathrm{c}}}p\left( E^{\prime }\right) dE^{\prime }=\frac{2}{3%
}\sqrt{\frac{2}{m_{\mathrm{n}}}}n\left( E^{\ast }\right) n_{\mathrm{pc}%
}\sigma _{\mathrm{m}}g_{-}\left( T\right) f_{-}\left( E^{\ast }\right) E_{%
\mathrm{c}}^{3/2}.  \label{p}
\end{equation}

Once converted, neutrons may get lost through the inverse process of
up-scattering with cross section $\propto g_{+}\left( T\right) $ that
depends on the deviation of the system from the magnetic ground state. The
corresponding spectral rate density is given by%
\begin{equation}
\tilde{p}\left( E\right) =\int_{0}^{\infty }\Phi \left( E\right) \Sigma
^{+}\left( E\rightarrow E^{\prime }\right) dE^{\prime }=n\left( E\right) n_{%
\mathrm{pc}}\sigma _{\mathrm{m}}g_{+}\left( T\right) f_{+}\left( E\right)
v\left( E+E^{\ast }\right) .  \label{p-welle(E)}
\end{equation}%
For the special case of UCN up-scattering, one obtains%
\begin{equation}
\tilde{p}=\int_{0}^{E_{\mathrm{c}}}\tilde{p}\left( E\right) dE=n_{\mathrm{UCN%
}}n_{\mathrm{pc}}\sigma _{\mathrm{m}}g_{+}\left( T\right) f_{+}\left(
0\right) v\left( E^{\ast }\right) ,  \label{p-welle}
\end{equation}%
where%
\begin{equation}
n_{\mathrm{UCN}}=\int_{0}^{E_{\mathrm{c}}}n\left( E\right) dE
\end{equation}%
is the UCN density. Notice that in the eqs.\ \ref{p(E')} and \ref{p-welle(E)}
the quantity multiplied with the spectral neutron density can be interpreted
as a corresponding rate constant. For conversion to neutrons with energy $%
E^{\prime }\ll E^{\ast }$ we write it as%
\begin{equation}
\tau _{\mathrm{conv}}^{-1}=n_{\mathrm{pc}}\sigma _{\mathrm{m}}g_{-}\left(
T\right) f_{-}\left( E^{\ast }\right) v\left( E^{\prime }\right) ,
\label{tau_UCN}
\end{equation}%
and the up-scattering rate constant as%
\begin{equation}
\tau _{\mathrm{up}}^{-1}=n_{\mathrm{pc}}\sigma _{\mathrm{m}}g_{+}\left(
T\right) f_{+}\left( 0\right) v\left( E^{\ast }\right) .  \label{tau_up}
\end{equation}%
Relevant is also the rate constant for neutron absorption, which is given by 
\begin{equation}
\tau _{\mathrm{a}}^{-1}=v\Sigma _{\mathrm{a}},  \label{tau_a}
\end{equation}%
where $\Sigma _{\mathrm{a}}$ is the macroscopic neutron absorption cross
section. In contrast to $\tau _{\mathrm{conv}}^{-1}$ and $\tau _{\mathrm{up}%
}^{-1}$, $\tau _{\mathrm{a}}^{-1}$ does not dependent on the neutron speed,
since $\Sigma _{\mathrm{a}}\propto 1/v$.

Turning now toward a discussion of realistic paramagnetic media, we start
with a material with a particularly low absorption, composed of the
paramagnetic atoms or molecules listed in Table $\ref{Species}$, implanted
in a matrix of $^{4}$He. A viable method employs injection of an impurity
loaded helium gas jet into superfluid helium, which makes a jelly-like
helium-impurity condensate \cite{Efimov/2013}. A high frequency discharge
prior to injection can lead to samples wherein impurity atoms are present
with a degree of dissociation up to $50$\%, and still $20$\% for samples
with highest atomic density \cite{Gordon/1988}. After injection, polarizable
heavy impurities can be packed in dry van der Waals clusters where the
impurity is surrounded by a crust of some dozen helium atoms. For molecular
nitrogen, impurity number densities up to $1.46\times 10^{21}\ \mathrm{cm}%
^{-3}$ were demonstrated \cite{Gordon/1993}, and one can expect oxygen to
behave similarly. An obvious obstacle in using the particularly weakly
absorbing $^{15}$N is the need for highly enriched material to suppress the
large absorption of $^{14}$N. This makes atomic $^{15}$N but also
paramagnetic nitric molecules such as $^{15}$NO ($S=1/2$) rather
impractical. Isotopically pure deuterium is commercially available in large
quantities but it remains to be shown experimentally that dry van der Waals
clusters can be produced with sufficient abundance of $^{2}$H atoms.
However, from a look in Table $\ref{Species}$ and also from a practical
point of view there seems to be no advantage compared to molecular oxygen
with natural isotopic composition.

In fact, molecular oxygen appears to be the most promising paramagnetic
species. No dissociation of molecules and subsequent stabilization of atoms
in the helium is needed here and, contrary to atomic Zeeman states, no
magnetic field needs to be applied for lifting degeneracies of magnetic
levels. The goal is to keep the O$_{2}$ molecules paramagnetic in a highly
packed state, avoiding magnetic order as it appears for instance in the
antiferromagnetic crystalline $\alpha $ phase of pure oxygen. For the
maximum density achieved in the aforementioned van der Waals solid the
molecules are already more than sufficiently separated. Indeed, neglecting
superexchange between O$_{2}$ molecules separated by $^{4}$He atoms and
using the parameters of the Heisenberg interaction given in \cite%
{Freiman/2004}, the spin interaction energy is found to be less than a 
\textrm{mK}, ensuring the paramagnetic state of such a system.

Even higher O$_{2}$ densities may prevail in clathrate hydrates, a special
class of inclusion compounds with water molecules forming an ice-like
hydrogen bonded network that contains sub-nm sized cavities stabilized by
guests of noble gas atoms or a wide range of molecules \cite{Sloan/2008}.
Methods of sample preparation of such systems from water ice can be found in
refs.\ \cite{Kuhs/1997,Kuhs/2000}. Oxygen molecules stabilize the type-II
clathrate hydrate structure that crystallizes in space group Fd\={3}m. Its
face centered cubic unit cell with size $a\approx 1.73\ \mathrm{nm}$
possesses $24$ cavities ($16$ with radius $0.395\ \mathrm{nm}$ and $8$ with
radius $0.473\ \mathrm{nm}$), corresponding to a cavity number density of $%
4.63\times 10^{21}\ \mathrm{cm}^{-3}$. For a filling fraction of $90$\%, the
cages are occupied only with single O$_{2}$ molecules, as was established in
a neutron diffraction study \cite{Chazallon/2002}. An inelastic neutron
scattering study on this system has revealed a dispersion-free magnetic
excitation with energy $0.4\ \mathrm{meV}$ \cite{Chazallon2/2002}. i.e.\
close to the zero-field splitting constant $D=0.491\ \mathrm{meV}$ measured
via ESR for gaseous oxygen \cite{Tinkham/1955}. Note that the O$_{2}$
density is still a factor five lower than in $\alpha $-oxygen. Using the
same argument as for the O$_{2}$-$^{4}$He van der Waals clusters one can
expect the system to stay paramagnetic still down to sub-Kelvin temperatures.

Molecular oxygen can also be intercalated in the fcc lattice of crystallized
C$_{60}$ molecules. Up to one O$_{2}$ molecule per C$_{60}$ unit can be
trapped on the octahedral sites, corresponding to a maximum molecular number
density of $1.38\times 10^{21}\ \mathrm{cm}^{-3}$. A neutron scattering
study performed on a system prepared with $70$\% site occupancy has
established a dispersion-free magnetic mode with energy $0.4\ \mathrm{meV}$ 
\cite{Renker/2001}, i.e.\ as observed in the O$_{2}$-hydrate clathrate. This
is a strong hint that in both cases the excitation is caused by the
zero-field splitting in the oxygen molecule, slightly shifted due to
environmental perturbation of the molecular Hamiltonian.

Table $\ref{Host medium}$ quotes, for the media discussed above, the rate
constants for neutron conversion to a typical UCN with a speed of $5\ 
\mathrm{m/s}$ and for absorption, according to eqs.\ \ref{tau_UCN} and \ref%
{tau_a}. While the van der Waals system has the lowest absorption, the
clathrate hydrate converts neutrons fastest. The intercalated C$_{60}$
system has the strongest absorption due to the large abundance of carbon
nuclei, and the smallest conversion. On the other hand, still stable at room
temperature, it might be the easiest to deal with. Rate constants for
neutron up-scattering (see eq.\ \ref{tau_up}) are shown in Fig.\ $2$, from
which one can read off values for the temperature $T_{=}$ where the
break-even with absorption occurs, i.e.\ $\tau _{\mathrm{up}}^{-1}=\tau _{%
\mathrm{a}}^{-1}$. They are also listed in Table $\ref{Host medium}$ (note
that for the C$_{60}$ system the absorption is too large for a break-even to
exist). Although, due to eq.\ \ref{g+}, one can always arrange for $\tau _{%
\mathrm{up}}^{-1}\ll \tau _{\mathrm{a}}^{-1}$, the intended technical
implementation of the moderator will set practical limits. The answer of the
question beyond which medium temperature the up-scattering becomes a
nuisance depends on the time the converted neutrons have to travel in the
material before escape, and hence on the neutron speed and the size of the
converter. Obviously, for a small converter the requirements are weaker than
for a big moderator discussed further below. 
\begin{table}[tbp] \centering%
\begin{tabular}{|l|l|l|l|l|l|}
\hline
Host & Cage structure & $n_{\mathrm{pc}}\text{ (}\mathrm{cm}^{-3}\text{)}$ & 
$\tau _{\mathrm{a}}^{-1}\text{ (}\mathrm{s}^{-1}\text{)}$ & $\tau _{\mathrm{%
UCN}}^{-1}\text{ (}\mathrm{s}^{-1}\text{)}$ & $T_{=}$ ($\mathrm{K}$) \\ 
\hline
$^{4}$He & $^{4}\text{He van der Waals clusters}$ & $1.46\times 10^{21}$ & $%
0.122$ & $3.36$ & $0.63$ \\ 
D$_{2}$O & $\text{type-II clathrate hydrate (90\%)}$ & $4.16\times 10^{21}$
& $7.43$ & $9.57$ & $1.09$ \\ 
carbon & $\text{hedragonal voids in fcc-C}_{60}$ (70\%) & $0.97\times
10^{21} $ & $63.9$ & $2.23$ & $-$ \\ \hline
\end{tabular}%
\caption{Rate constants for neutron absorption and for production of UCN with $5$ m/s, in three media 
hosting isolated O$_{2}$ molecules at low temperature ($g_{-}=4/3$). All values are given for bulk matter 
(for the clathrate and the intercalated fcc-C$_{60}$ with the indicated O$_{2}$ filling fractions of cages). At
the medium temperature $T_{=}$ the rate constants for up-scattering and absorption are equal.}%
\label{Host medium}%
\end{table}%

\begin{figure}[tbp]
\centering
\includegraphics[width=0.8\textwidth]{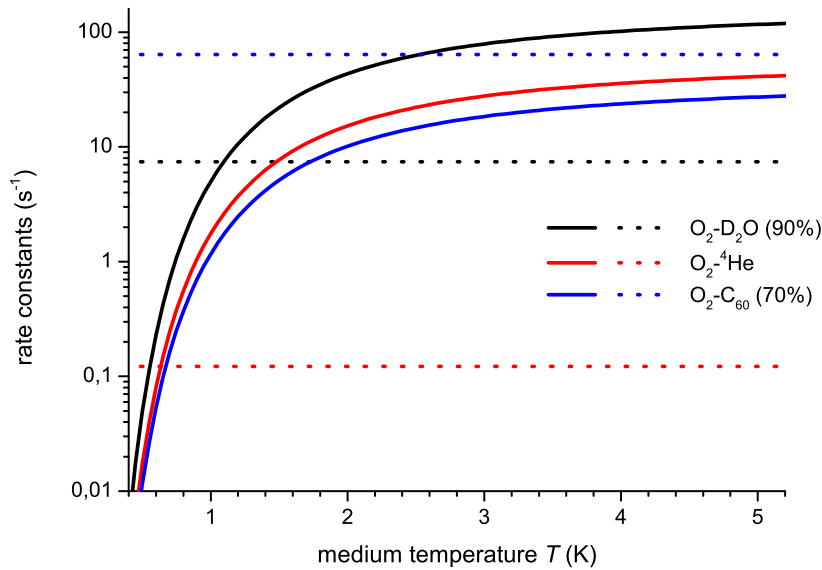}
\caption{Rate constants $\protect\tau _{%
\mathrm{up}}^{-1}$ for neutron up-scattering (solid lines) and $\protect\tau _{\mathrm{a}}^{-1}$ for neutron absorption
(dotted) for the three paramagnetic inclusion compounds of molecular oxygen
listed in Table \protect\ref{Host medium}.}
\label{fig:figure2}
\end{figure}

For a comparison of neutron conversion rates in different materials it is
useful to refer to an incident Maxwellian neutron spectrum characterized by
a temperature $T_{\mathrm{n}}$ and with density given by%
\begin{equation}
n\left( E,T_{\mathrm{n}}\right) dE=\frac{2}{\sqrt{\pi }}\frac{n}{\left( k_{%
\mathrm{B}}T_{\mathrm{n}}\right) ^{3/2}}\exp \left( -\frac{E}{k_{\mathrm{B}%
}T_{\mathrm{n}}}\right) \sqrt{E}dE,  \label{n(E,T_n)}
\end{equation}%
where%
\begin{equation}
n=\int_{0}^{\infty }n\left( E,T_{\mathrm{n}}\right) dE
\end{equation}%
is the total neutron density \cite{Beckurts/1964}. It is related to the
total flux by%
\begin{equation}
\Phi =\frac{2}{\sqrt{\pi }}n\sqrt{\frac{2k_{\mathrm{B}}T_{\mathrm{n}}}{m_{%
\mathrm{n}}}}.  \label{total flux}
\end{equation}%
A well understood medium for neutron conversion is pure superfluid $^{4}$He
at saturated vapor pressure. It was first analysed by Pendlebury \cite%
{Pendlebury/1982}, and we use it as a benchmark here. Neutrons may become
converted to UCN if they have an energy of about $E^{\ast }\approx 1\ 
\mathrm{meV}$ (corresponding to wavenumber $k^{\ast }\approx 7\ \mathrm{nm}%
^{-1}$) where the neutron and helium dispersion curves cross each other.
Neglecting a small contribution due to multi-phonon processes \cite%
{Schmidt-Wellenburg/2009, Baker/2003,Korobkina/2002}, the UCN conversion
rate follows from eq.\ $3.37$ in ref.\ \cite{Golub/1991}, which after
integration over energy can be written as%
\begin{equation}
p_{\mathrm{He}}=\frac{2}{3}\Phi n_{\mathrm{He}}\sigma _{\mathrm{He}}S\left(
k^{\ast }\right) \alpha E_{\mathrm{c}}^{3/2}\exp \left( -\frac{E^{\ast }}{k_{%
\mathrm{B}}T_{\mathrm{n}}}\right) \frac{\sqrt{E^{\ast }}}{\left( k_{\mathrm{B%
}}T_{\mathrm{n}}\right) ^{2}}  \label{p-He}
\end{equation}%
where $n_{\mathrm{He}}\approx 2.18\times 10^{22}\ \mathrm{cm}^{-3}$ is the $%
^{4}$He atom density, $\sigma _{\mathrm{He}}\approx 1.34\ \mathrm{barn}$ is
the coherent scattering cross section per helium atom, $S\left( k^{\ast
}\right) \approx 0.105$ is the static structure factor of superfluid $^{4}$%
He at $k^{\ast }$ \cite{Cowley/1971}, and $\alpha \approx 1.45$ accounts for
the overlap of the two dispersion curves. The factor $E_{\mathrm{c}}^{3/2}$
is the same as appearing in the similar expression given in eq.\ \ref{p}.
Figure $3$ shows that, despite $n_{\mathrm{He}}/n_{\mathrm{pc}}\approx 5$
and for any temperatures of practical interest, the fully deuterated O$_{2}$%
-hydrate with $90$\% cage occupancy has a higher UCN conversion rate than
superfluid $^{4}$He, which is mainly due to the absence of the unfavorable
structure factor $S\left( k^{\ast }\right) $. For $T_{\mathrm{n}}=100\ 
\mathrm{K}$ ($T_{\mathrm{n}}=30\ \mathrm{K}$), $p/p_{\mathrm{He}}=2.9\ $($%
3.4 $). At neutron temperature close to $6\ \mathrm{K}$, where conversion in
superfluid $^{4}$He works best, the clathrate converts neutrons a factor $9$
better. The highest conversion in the clathrate appears at an optimum
neutron temperature about $2.3\ \mathrm{K}$. Note that the cooling cascade
discussed in the next section may enhance production of UCN and VCN in a
paramagnetic medium by a further and even larger factor. Note also that the
analysis has taken into account only the paramagnetic cross section,
neglecting any contribution due to phonons.

\begin{figure}[tbp]
\centering
\includegraphics[width=0.8\textwidth]{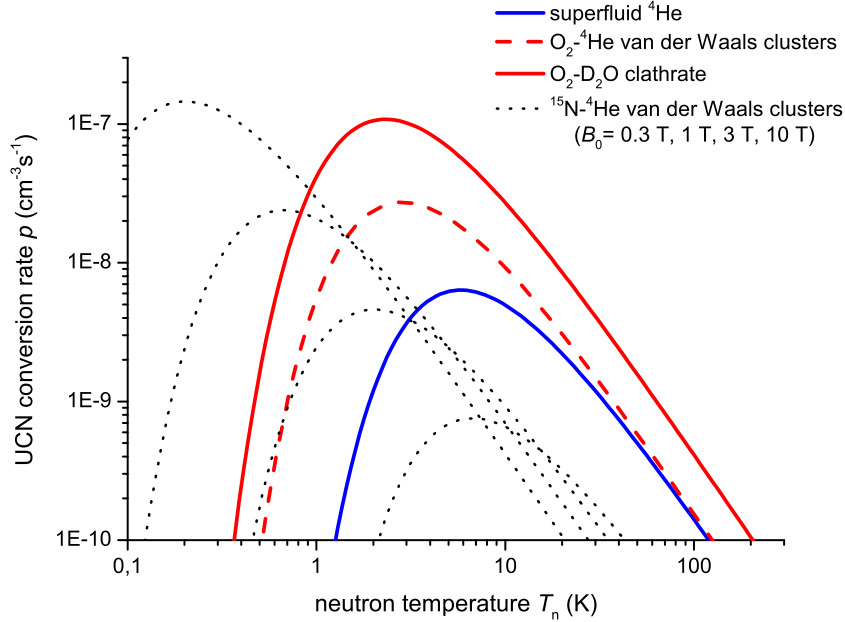}
\caption{Rates of neutron conversion to UCN with energy up to $E_{\mathrm{c}%
}=250\ \mathrm{neV}$, calculated using eq.\ \protect\ref{p} in the
low-temperature limit with parameters for the various systems as given in
Table \protect\ref{Host medium} (we set $E^{\ast }=0.49\ \mathrm{meV}$ for
the O$_{2}$-$^{4}$He van der Waals clusters), and eq.\ \protect\ref{p-He}
for superfluid $^{4}$He. The neutron flux is set to $\Phi =1\ \mathrm{cm}%
^{-2}\mathrm{s}^{-1}$. Calculations for the $^{15}$N-$^{4}$He van der Waals
clusters are done for $1.8\times 10^{20}\ \mathrm{cm}^{-3}$ atomic density
of $^{15}$N (i.e.\ the maximum value reported in \protect\cite{Gordon/1988}%
), for $f_{-}=1$, and for values of an external magnetic field $B_{0}$ as
quoted and corresponding to the dotted curves from left to right.}
\label{fig:figure3}
\end{figure}

\section{Paramagnetic cascade cooling of neutrons}

For analysis of cascade cooling it is useful to consider first an infinite
medium, for which the neutron transport equation takes a particularly simple
form. For the sake of a transparent analytical treatment, we consider only
the energy transfers in units of $E^{\ast }$ due to paramagnetic scattering,
neglecting any other inelastic channels and hence underestimating the true
moderation efficiency of the material.

We define groups $j$ of neutrons characterized by a spatial density $%
n_{j,\Delta }dE$ in an energy interval $dE$ about%
\begin{equation}
E_{j}=jE^{\ast }+\Delta  \label{E_j}
\end{equation}%
where $0\leq \Delta <E^{\ast }$ defines a base energy for the lowest group, $%
j=0$. For simplicity of notation we will omit the index $\Delta $ in the
sequel. Note that, in a $j$-changing scattering process involving
non-dispersive excitiations, $dE$ does not change. We denote the rate
constant for scattering from group $j$ to $j^{\prime }$ by $\tau
_{j\rightarrow j^{\prime }}^{-1}$. The rate equation for the population of
the group $j$ can then be written as 
\begin{equation}
\frac{dn_{j}}{dt}=s_{j}+n_{j+1}\tau _{j+1\rightarrow j}^{-1}+n_{j-1}\tau
_{j-1\rightarrow j}^{-1}-n_{j}\tau _{j\rightarrow j-1}^{-1}-n_{j}\tau
_{j\rightarrow j+1}^{-1}-n_{j}\tau _{\mathrm{a}}^{-1}.  \label{transport}
\end{equation}%
The term $s_{j}$ describes homogeneously distributed sources of neutrons.
The second and third terms describe feeding due to down-scatters from the
group $j+1$ and due to up-scatters from the group $j-1$. The fourth and
fifth terms describe losses due to down-scatters to the group $j-1$ and due
to up-scatters to the group $j+1$. The last term describes absorption losses
with rate constant $\tau _{\mathrm{a}}^{-1}$, which is universal for all
groups according to eq.\ \ref{tau_a} and $\Sigma _{\mathrm{a}}\propto 1/v$.
The system of first-order differential equations\ \ref{transport} can be
written as 
\begin{equation}
\frac{dn_{j}}{dt}=M_{jk}n_{k}+s_{j},
\end{equation}%
with a tridiagonal matrix%
\begin{equation}
\mathbf{M}=\left( 
\begin{array}{ccccc}
-\tau _{0\rightarrow 1}^{-1}-\tau _{\mathrm{a}}^{-1} & \tau _{1\rightarrow
0}^{-1} & 0 & 0 &  \\ 
\tau _{0\rightarrow 1}^{-1} & -\tau _{1\rightarrow 0}^{-1}-\tau
_{1\rightarrow 2}^{-1}-\tau _{\mathrm{a}}^{-1} & \tau _{2\rightarrow 1}^{-1}
& 0 &  \\ 
0 & \tau _{1\rightarrow 2}^{-1} & -\tau _{2\rightarrow 1}^{-1}-\tau
_{2\rightarrow 3}^{-1}-\tau _{\mathrm{a}}^{-1} & \tau _{3\rightarrow 2}^{-1}
&  \\ 
0 & 0 & \tau _{2\rightarrow 3}^{-1} & -\tau _{3\rightarrow 2}^{-1}-\tau
_{3\rightarrow 4}^{-1}-\tau _{\mathrm{a}}^{-1} &  \\ 
&  & \cdots &  & \ddots%
\end{array}%
\right)  \label{M}
\end{equation}%
of constant coefficients (the value of $\Delta $ is kept fixed). Stationary
solutions are given by%
\begin{equation}
n_{j}=-\left( \mathbf{M}^{-1}\right) _{jk}s_{k}.  \label{n_j}
\end{equation}%
We assume that the sources emit neutrons with a Maxwellian spectrum as given
in eq.\ \ref{n(E,T_n)} and define accordingly%
\begin{equation}
s_{j}=\frac{2}{\sqrt{\pi }}\frac{n}{\tau }\frac{\sqrt{E_{j}}}{\left( k_{%
\mathrm{B}}T_{\mathrm{n}}\right) ^{3/2}}\exp \left( -\frac{E_{j}}{k_{\mathrm{%
B}}T_{\mathrm{n}}}\right) .  \label{s_j}
\end{equation}%
The source strength is characterized by the density rate $n\tau ^{-1}$. The
rate constants for neutron up- and down-scattering from the group $j$ follow
from%
\begin{equation}
\tau _{j\rightarrow j\pm 1}^{-1}=v_{j}\int_{0}^{\infty }\Sigma ^{\pm }\left(
E\rightarrow E_{j\pm 1}\right) dE,
\end{equation}%
where%
\begin{equation}
v_{j}=\sqrt{\frac{2E_{j}}{m_{\mathrm{n}}}}
\end{equation}%
is the speed of neutrons in group $j$, and $\Sigma ^{\pm }$ is the
macroscopic inelastic scattering cross section from eq.\ \ref{Sigma+---+}.
One obtains%
\begin{equation}
\tau _{j\rightarrow j\pm 1}^{-1}=n_{\mathrm{pc}}\sigma _{\mathrm{m}}g_{\pm
}\left( T\right) f_{\pm }\left( E_{j}\right) v_{j\pm 1}.
\end{equation}

Solving the system of linear equations\ \ref{n_j} requires inversion of the
square matrix $\mathbf{M}$ which can be done only for finite matrix order $l$%
. For a source spectrum as given in eq.\ \ref{s_j} and for neutron group
number $j<l$, not too close to $l$, values for $n_{j}$ do converge when
calculated using matrices with increasing order. In practice one chooses $l$
large enough to cover a major part of the source spectrum. For instance, $%
l=100$ is more than sufficient for $T_{\mathrm{n}}\leq $ $30\ \mathrm{K}$.
Results of calculations performed that way are shown in Fig.\ $4$, for the
fully deuterated O$_{2}$-clathrate hydrate held at various temperatures. One
can see that, the colder the medium, the more the paramagnetic cooling
cascade compresses neutron spectra toward lower neutron energy. One observes
also a large enhancement of the group-$0$ density (here exemplarily shown
for UCN with energy $100\ \mathrm{neV}$) by more than two orders of
magnitude, with respect to the situation of thermal equilibrium between the
moderator and the neutron sources at $T_{\mathrm{n}}=30\ \mathrm{K}$. Figure 
$5$ presents stationary neutron densities in the energy region of UCN and
VCN, obtained by variation of the offset energy $\Delta $ in group $0$.
Figures $6$ and $7$ show neutron density spectra calculated for the van der
Waals cluster system with an O$_{2}$ density as quoted in Table \ref{Host
medium} and otherwise the same parameters as used in Figs.\ $4$ and $5$. The
comparison of the two media complies with the expectation that lower
absorption leads to larger neutron densities in the low-energy groups and an
increase of the ratio $n_{0}/n_{1}$ due to an improved UCN accumulation in
the medium. One can also see that for lower absorption more neutron groups
contribute effectively to the cooling as the higher groups get stronger
depleted.

\begin{figure}[tbp]
\centering
\includegraphics[width=0.8\textwidth]{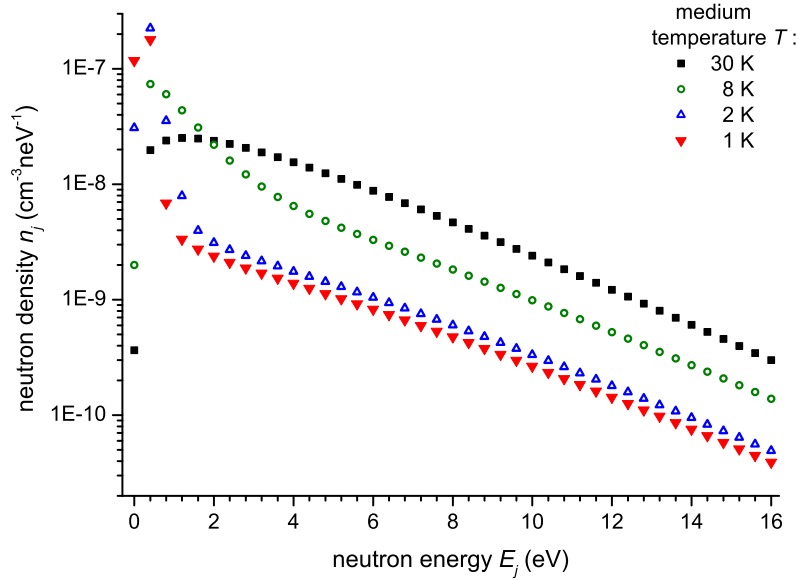}
\caption{Stationary spectral
neutron densities in fully deuterated O$_{2}$-clathrate hydrate with $90$\%
cage occupancy, for $T_{\mathrm{n}}=30\ \mathrm{K}$ and source strength $n%
\protect\tau ^{-1}=1\ \mathrm{cm}^{-3}\mathrm{s}^{-1}$ (see eq.\ \protect\ref%
{s_j}). Each point belongs to a neutron group with energy $jE^{\ast }+\Delta 
$ (shown for $\Delta =100\ \mathrm{neV}$).}
\label{fig:figure4}
\end{figure}

\begin{figure}[tbp]
\centering
\includegraphics[width=0.8\textwidth]{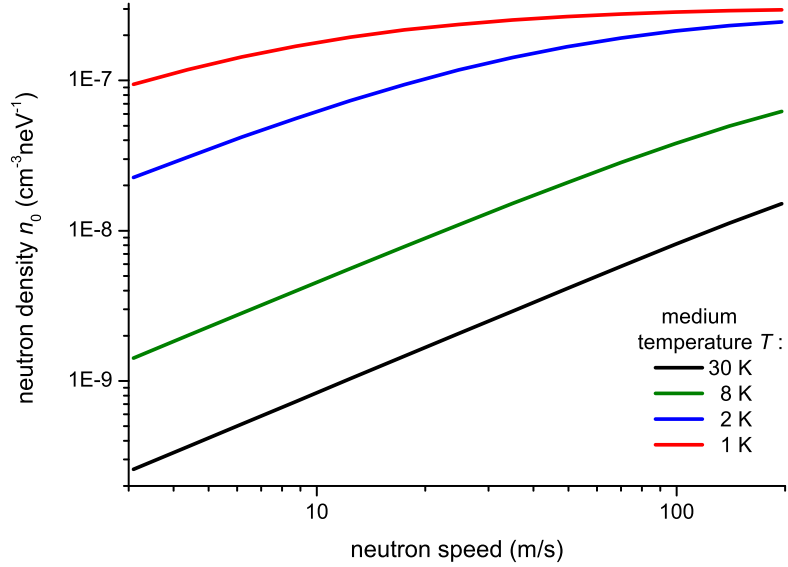}
\caption{Stationary spectral
neutron densities in the lowest neutron group for the fully deuterated O$%
_{2} $-clathrate hydrate with parameters as in Fig.\ $4$ but energies $%
\Delta $ in the region of UCN and VCN. The range of neutron speed
corresponds to $5\times 10^{-8}\ \mathrm{neV}\leq \Delta \leq 2\times
10^{-4}\ \mathrm{neV}$.}
\label{fig:figure5}
\end{figure}

\begin{figure}[tbp]
\centering
\includegraphics[width=0.8\textwidth]{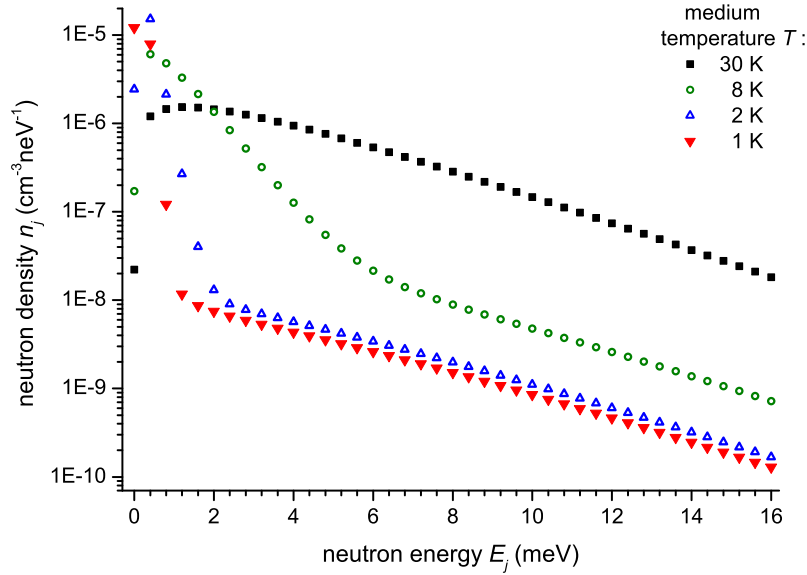}
\caption{Stationary spectral neutron densities in
the O$_{2}$-$^{4}$He van der Waals cluster system with O$_{2}$ density $%
1.46\times 10^{21}\ \mathrm{cm}^{-3}$, for $T_{\mathrm{n}}=30\ \mathrm{K}$
and source strength $n\protect\tau ^{-1}=1\ \mathrm{cm}^{-3}\mathrm{s}^{-1}$
(see eq.\ \protect\ref{s_j}). Each point belongs to a neutron group with
energy $jE^{\ast }+\Delta $ (shown for $\Delta =100\ \mathrm{neV}$).}
\label{fig:figure6}
\end{figure}

\begin{figure}[tbp]
\centering
\includegraphics[width=0.8\textwidth]{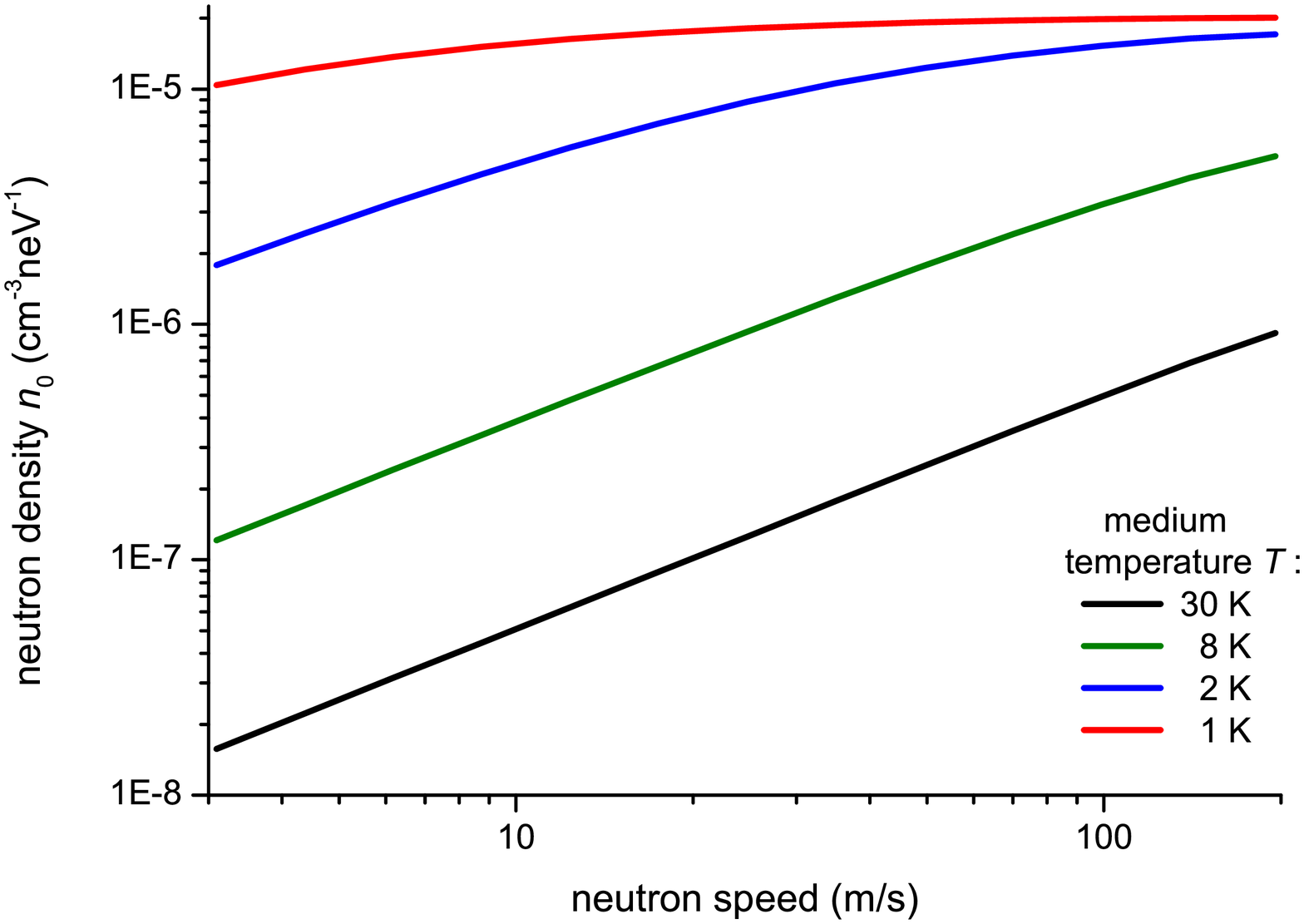}
\caption{Stationary spectral neutron densities in the lowest neutron group for
the O$_{2}$-$^{4}$He van der Waals system with parameters as in Fig.\ $6$
but energies $\Delta $ in the region of UCN and VCN. The range of neutron
speed corresponds to $5\times 10^{-8}\ \mathrm{neV}\leq \Delta \leq 2\times
10^{-4}\ \mathrm{neV}$.}
\label{fig:figure7}
\end{figure}

It is also interesting to see how the source spectrum temperature $T_{%
\mathrm{n}}$ influences the shape of the neutron density spectrum in a cold
moderator, and in particular its component $n_{0}$. Figure $8$ presents
examples for the O$_{2}$-clathrate hydrate. The values $T_{\mathrm{n}}=$ $%
30\ \mathrm{K}$ and $100\ \mathrm{K}$ are representative for Maxwellians as
frequently employed to approximate the (usually undermoderated) neutron
spectra from liquid deuterium or liquid hydrogen cold sources in
superpositions with similar weights. The three curves for $T=T_{\mathrm{n}}$
represent spectra of the neutron sources, noting that under this condition
the moderator has no influence on the spectral shape. One finds that at $%
T=T_{\mathrm{n}}=100\ \mathrm{K}$ ($300\ \mathrm{K}$) the density $n_{0}$
(again exemplarily taken as UCN with energy $100\ \mathrm{neV}$) is lower
than at $T=T_{\mathrm{n}}=30\ \mathrm{K}$ by a factor of $6$ ($32$), whereas
for a cold moderator held at $T=1\ \mathrm{K}$, $n_{0}$ decreases by only a
factor $1.3$ ($2.8$) if $T_{\mathrm{n}}=100\ \mathrm{K}$ ($300\ \mathrm{K}$)
instead of $30\ \mathrm{K}$. These numbers tell us that precooling of
neutrons by a liquid deuterium or liquid hydrogen cold source is sufficient
for the paramagnetic cooling cascade to reach almost its full performance.
Direct paramagnetic cooling of thermal neutrons on the other hand involves
much longer cascades and suffers from a suppression of the inelastic
scattering cross section for large neutron energies due to the magnetic form
factor (compare Fig.\ $1$). However, the present analysis still neglects the
experimentally studied, non-magnetic excitations in clathrate hydrates \cite%
{Chazallon2/2002}, which are able to remove many meV of kinetic energy from
the neutron in single scattering events. They might shortcut many steps of
the paramagnetic cooling process and thereby provide an intrinsic precooling
to a degree that a separate cold source could become unnecessary.

\begin{figure}[tbp]
\centering
\includegraphics[width=0.8\textwidth]{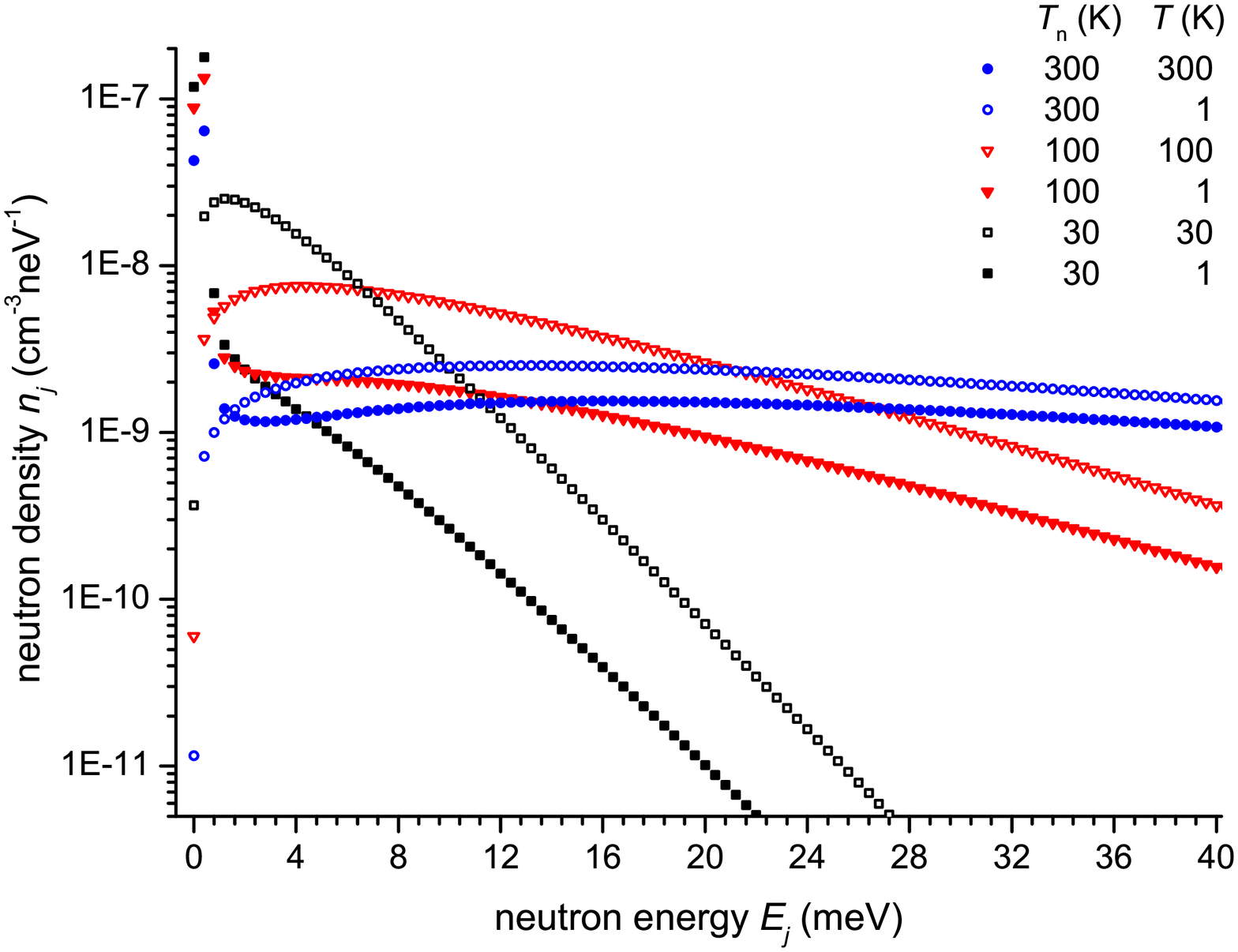}
\caption{Stationary spectral neutron densities in the
fully deuterated O$_{2}$-clathrate hydrate with $90$\% cage occupancy, for
various pairs of temperatures $\left( T_{\mathrm{n}},T\right) $ and source
strength $n\protect\tau ^{-1}=1\ \mathrm{cm}^{-3}\mathrm{s}^{-1}$ (see eq.\ 
\protect\ref{s_j}). Each point belongs to a neutron group with energy $%
jE^{\ast }+\Delta $ (shown for $\Delta =100\ \mathrm{neV}$).}
\label{fig:figure8}
\end{figure}

From the previous discussion it is qualitatively clear that, the higher $T_{%
\mathrm{n}}$, the stronger the relative contribution of the higher-energy
neutron groups to the moderated density $n_{0}$. This can indeed be
quantified by solving eq.\ \ref{n_j} for a system limited to $l$ neutron
groups and considering the $l$ dependence of $n_{0}$. The matrix operating
on the system of groups, $j=0$ to $j=l-1$, has to be properly defined,
ensuring the absence of transitions to or from groups with larger $j$. While
simple truncation of $\mathbf{M}$ from eq.\ \ref{M} to order $l$ removes
feeding from groups $j\geq l$, losses to groups $j\geq l$ are avoided by
removing the rate constant $\tau _{l\rightarrow l+1}^{-1}$ from the element $%
M_{ll}$. We denote the correspondingly modified matrix by $\mathbf{M}%
_{l\times l}$. For instance,%
\begin{equation}
\mathbf{M}_{2\times 2}=\left( 
\begin{array}{cc}
-\tau _{0\rightarrow 1}^{-1}-\tau _{\mathrm{a}}^{-1} & \tau _{1\rightarrow
0}^{-1} \\ 
\tau _{0\rightarrow 1}^{-1} & -\tau _{1\rightarrow 0}^{-1}-\tau _{\mathrm{a}%
}^{-1}%
\end{array}%
\right)
\end{equation}%
connects only the two lowest neutron groups, $j=0$ and $1$, a situation
reminiscent of neutron conversion.

\begin{figure}[tbp]
\centering
\includegraphics[width=0.8\textwidth]{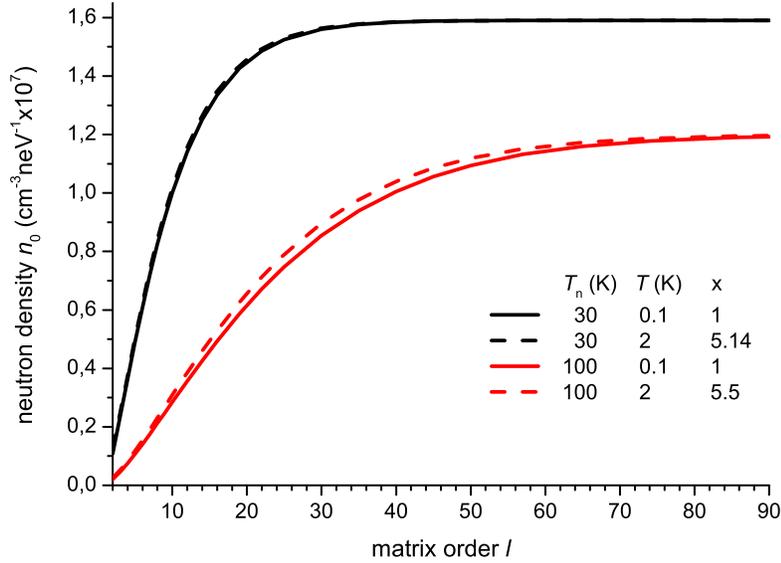}
\caption{Stationary spectral UCN density (for $\Delta =100\ \mathrm{neV}$) in the
fully deuterated O$_{2}$-clathrate hydrate with $90$\% cage occupancy, as a
function of the order $l$ of the matrix $\mathbf{M}_{l\times l}$. The source
strength is $n\protect\tau ^{-1}=1\ \mathrm{cm}^{-3}\mathrm{s}^{-1}$ (see
eq.\ \protect\ref{s_j}). Results shown for $T=2\ \mathrm{K}$ are normalized
to the saturation level at $T=0.1\ \mathrm{K}$ by the respective factor $x$.}
\label{fig:figure9}
\end{figure}

Figure $9$ shows densities $n_{0}$ as a function of the matrix order $l$,
for the fully deuterated O$_{2}$-clathrate hydrate. One observes only small
differences in the $l$ dependence for $T=0.1\ \mathrm{K}$ and $2\ \mathrm{K}$%
. Higher temperatures are less interesting if one wants to take advantage of
the extraordinary thermal conductance of superfluid helium as a cooling
agent for the clathrate (requiring $T<2.17\ \mathrm{K}$ beyond which the
helium becomes a normal liquid at saturated vapor pressure). One can see
that for a cold neutron source spectrum at $30\ \mathrm{K}$ already eight
groups are sufficient to attain half of the saturation density, whereas for $%
T_{\mathrm{n}}=100\ \mathrm{K}$ about twenty groups are needed. The number
of groups one wants to participate in the cooling process has impact on the
necessary size of a real moderator which is discussed further below.

The effect of the complete cooling cascade can be deduced from comparing the
group-$0$ densities in a large system of groups and in the two-group system
with matrix $\mathbf{M}_{2\times 2}$. We define correspondingly a cascade
enhancement factor%
\begin{equation}
\eta _{\mathrm{cascade}}\left( T,T_{\mathrm{n}}\right) =\frac{n_{0}\left(
T,T_{\mathrm{n}},l\rightarrow \infty \right) }{n_{0}\left( T,T_{\mathrm{n}%
},l=2\right) }.  \label{eta-cascade}
\end{equation}%
It tells us how much the feeding from the groups $j>1$ improves the group-$0$
density. Figure $10$ shows cascade enhancements for different temperatures
and media. In thermal equilibrium between the moderator and the neutron
sources,%
\begin{equation}
n_{0}\left( T=T_{\mathrm{n}},l=2\right) =n_{0}\left( T=T_{\mathrm{n}%
},l\rightarrow \infty \right)
\end{equation}%
as they have to fulfill for a well defined $\mathbf{M}_{l\times l}$.
Correspondingly, $\eta _{\mathrm{cascade}}\left( T=T_{\mathrm{n}}\right) =1$%
. Note that the cascade enhancement accounts only for a part of the effects
visible in the figures before. Indeed, the total enhancement of $n_{0}$
observed when reducing the temperature $T$ below $T_{\mathrm{n}}$ can be
written as%
\begin{equation}
\eta =\eta _{\mathrm{cascade}}\left( T,T_{\mathrm{n}}\right) \eta _{\mathrm{%
conv}}\left( T,T_{\mathrm{n}}\right) ,
\end{equation}%
with the factor%
\begin{equation}
\eta _{\mathrm{conv}}\left( T,T_{\mathrm{n}}\right) =\frac{n_{0}\left( T,T_{%
\mathrm{n}},l=2\right) }{n_{0}\left( T=T_{\mathrm{n}},l=2\right) }
\end{equation}%
accounting for the temperature dependence of the medium as a converter.
Lowering $T$ leads to $\eta _{\mathrm{conv}}\left( T,T_{\mathrm{n}}\right)
>1 $ for two reasons. On one hand neutron up-scattering becomes suppresses
due to the factor $g_{+}\left( T\right) $ in the cross section. On the other
hand, due to its proportionality to $g_{-}\left( T\right) $, the
down-scattering cross section increases with the population of the magnetic
ground state. For O$_{2}$ for instance, $g_{-}\left( 1\ \mathrm{K}\right)
/g_{-}\left( 30\ \mathrm{K}\right) \approx \allowbreak 2.7$.

\begin{figure}[tbp]
\centering
\includegraphics[width=0.8\textwidth]{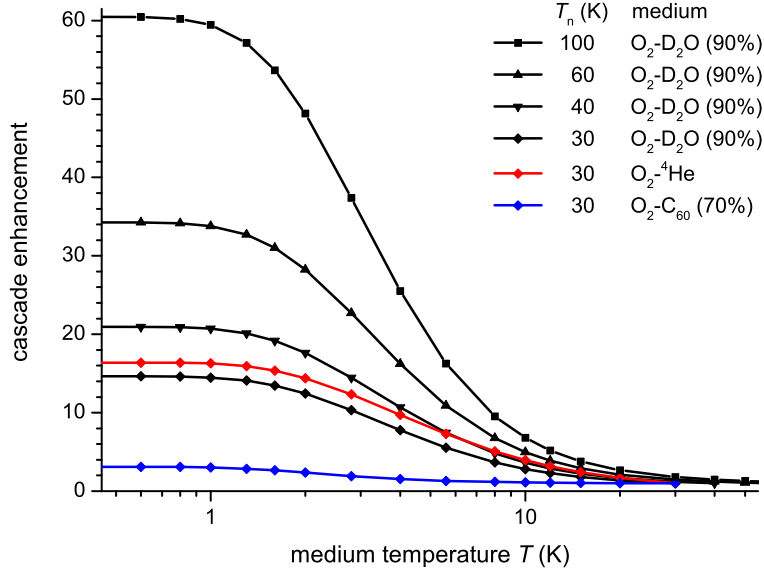}
\caption{Cascade enhancements as defined in eq.\ \protect
\ref{eta-cascade}, for the paramagnetic media quoted in Table \protect\ref%
{Host medium}.}
\label{fig:figure10}
\end{figure}

Finally, we give an estimate for the size a realistic moderator needs to
have for the cooling cascade to take effect, noting that the calculations
above were performed for an infinite medium. We consider the O$_{2}$%
-clathrate with $90$\% cage occupancy held at a temperature low enough that
we can neglect up-scattering. While a highly packed state of the clathrate
grains can be achieved using a press after clathration, some porosity of the
medium will however be useful for cooling the grains with superfluid helium.
In the sequel we take a packing fraction $\phi =0.75$ as a practical value.
Relevant quantities for this discussion are the macroscopic cross sections $%
\Sigma _{\mathrm{ie}}$, $\Sigma _{\mathrm{e}}$ and $\Sigma _{\mathrm{a}}$
for inelastic and elastic scattering for the neutron groups $j\geq 2$, and
for neutron absorption. An estimate for $\Sigma _{\mathrm{ie}}$ follows from
eq.\ \ref{Sigma+---+},%
\begin{equation}
\Sigma _{\mathrm{ie}}=n_{\mathrm{pc}}\sigma _{\mathrm{m}}\frac{k^{\prime }}{k%
}g_{-}\left( T\right) f_{-}\left( E\right) \approx 0.0076\ \mathrm{cm}^{-1},
\end{equation}%
where the value on the right side holds for $g_{-}\left( T\right) =4/3$ and $%
f_{-}\left( E\right) k^{\prime }/k=0.5$. Note that this value underestimates
the real cross section for the first seven neutron groups. The elastic cross
section $\Sigma _{\mathrm{e}}$ is mainly due to coherent scattering by the
packed clathrate crystallites with a contribution due to spin incoherent
scattering by the deuterons. Due to the large size of the fcc elementary
cell of the O$_{2}$-clathrate hydrate, the Bragg cutoff wavelength is $%
\lambda _{\mathrm{B}}\approx 2.0\ \mathrm{nm}$. Since neutrons with energy $%
E^{\ast }=0.4\ \mathrm{meV}$ have a wavelength of $1.41\ \mathrm{nm}$, any
neutron, prior to its final conversion into a UCN or VCN or absorption, will
diffuse through the medium. The cross section has a complex energy
dependence due to the Bragg edges. For an estimate we use (see, e.g., ref.\ 
\cite{Gurevich/1968}) an average value given by%
\begin{equation}
\Sigma _{\mathrm{e}}\approx n\left( \mathrm{O}\right) \sigma _{\mathrm{s}%
}\left( \mathrm{O}\right) +n\left( \mathrm{D}\right) \sigma _{\mathrm{s}%
}\left( \mathrm{D}\right) =0.41\ \mathrm{cm}^{-1},
\end{equation}%
where $n\left( \mathrm{O}\right) $ and $n\left( \mathrm{D}\right) $ are
macroscopic averages of atomic number densities of oxygen and deuterium, and 
$\sigma _{\mathrm{s}}\left( \mathrm{O}\right) \approx 4.23\ \mathrm{barn}$
and $\sigma _{\mathrm{s}}\left( \mathrm{D}\right) \approx 7.64\ \mathrm{barn}
$ are the total scattering cross sections per oxygen and deuterium atom. The
macroscopic absorption cross section in neutron group $j$ is given by%
\begin{equation}
\Sigma _{\mathrm{a}}=\frac{2200}{277\sqrt{j}}\times \left[ n\left( \mathrm{O}%
\right) \sigma _{\mathrm{a}}\left( \mathrm{O}\right) +n\left( \mathrm{D}%
\right) \sigma _{\mathrm{a}}\left( \mathrm{D}\right) \right] \leq 0.00014\ 
\mathrm{cm}^{-1},
\end{equation}%
where $\sigma _{\mathrm{a}}\left( \mathrm{O}\right) $ and $\sigma _{\mathrm{a%
}}\left( \mathrm{D}\right) $ are the atomic absorption cross sections of
oxygen and deuterium for neutrons with a speed of $2200\ \mathrm{m/s}$ (see
values in Table \ref{Species}), and the value on the right is for $j\geq 2$.
The cross sections are thus hierarchically ordered as%
\begin{equation}
\Sigma _{\mathrm{a}}\ll \Sigma _{\mathrm{ie}}\ll \Sigma _{\mathrm{e}}.
\end{equation}

For multiple inelastic scattering events of the cooling cascade to take
place, the moderator needs to be sufficiently large. In analogy to the
diffusion length with respect to neutron absorption \cite{Beckurts/1964}, we
define here a quantity 
\begin{equation}
L_{\mathrm{d}}=1/\sqrt{3\Sigma _{\mathrm{ie}}\left( \Sigma _{\mathrm{e}%
}+\Sigma _{\mathrm{ie}}\right) }.
\end{equation}%
It specifies the mean distance $\bar{r}$ between two inelastic scattering
events in presence of strong elastic diffusion,%
\begin{equation}
\bar{r}=2L_{\mathrm{d}}.
\end{equation}%
For the cross section estimates quoted above one finds%
\begin{equation}
L_{\mathrm{d}}\approx 10\ \mathrm{cm},
\end{equation}%
so that a fully deuterated O$_{2}$-clathrate hydrate moderator with linear
extension $D_{\func{mod}}$ less than a meter should provide efficient
paramagnetic cascade cooling. Due to up-scattering and absorption (see eqs.\ %
\ref{tau_up} and \ref{tau_a}), $D_{\func{mod}}$ defines a minimum speed that
the slowed down, very cold neutrons need to have for escaping from deep
inside the moderator,%
\begin{equation}
v_{\mathrm{VCN}}\gtrsim D_{\func{mod}}\tau _{\mathrm{VCN}}^{-1},
\label{v-VCN condition}
\end{equation}%
with%
\begin{equation}
\tau _{\mathrm{VCN}}^{-1}=\tau _{\mathrm{a}}^{-1}+\tau _{\mathrm{up}%
}^{-1}\left( T\right) .
\end{equation}%
The rate constants for both loss channels are independent on the neutron
speed $v_{\mathrm{VCN}}$ (for VCN this statement holds for $E_{\mathrm{VCN}%
}\ll E^{\ast }$) but $\tau _{\mathrm{up}}^{-1}$ depends on the moderator
temperature. A minimum necessary requirement is due to the fact that $\tau _{%
\mathrm{up}}^{-1}$ can always be suppressed below $\tau _{\mathrm{a}}^{-1}$
by choosing $T$ sufficiently low (see Fig.\ $2$ for break-even temperature
values for the media discussed before). For an O$_{2}$-clathrate moderator
with $D_{\func{mod}}\approx 1\ \mathrm{m}$ one concludes that, even at
lowest $T$, $v_{\mathrm{VCN}}\gtrsim 7.5\ \mathrm{m/s}$, which is the
highest neutron speed in a UCN spectrum defined by a high-potential wall
material. Turning around the argument leading to eq.\ \ref{v-VCN condition},
neutrons with lower speed will escape the medium only from within a certain
depth from the moderator surface, which may be further reduced if the group-$%
0$ neutrons are too strongly diffused \cite{Steyerl/1974}. For the moderator
held at $0.8\ \mathrm{K}$ ($1.2\ \mathrm{K}$, $1.5\ \mathrm{K}$, $2\ \mathrm{%
K}$) the minimum neutron speed as defined by eq.\ \ref{v-VCN condition}
increases to $9\ \mathrm{m/s}$ ($18.1\ \mathrm{m/s}$, $29.4\ \mathrm{m/s}$, $%
50.9\ \mathrm{m/s}$). Therefore, we may conclude that the fully deuterated O$%
_{2}$-clathrate hydrate moderator will be best suited for production of VCN.
These may either be used directly, e.g., for an advanced neutron-antineutron
search \cite{Young/2014} or for various other applications mentioned in the
introduction, or be transformed to UCN via gravity and/or a neutron turbine
as in ILL's long-standing UCN source.

\section{Conclusions}

This paper has presented a new mechanism for cooling neutrons well below
temperatures attained in liquid hydrogen and deuterium cold sources. Based
on the dispersion-free inelastic scattering in a paramagnetic material,
neutrons lose kinetic energy in constant steps $E^{\ast }$ defined by
electronic Zeeman energy or zero-field splittings of molecular magnetic
levels. The analytical expressions derived here reveal large possible gains
in the production of VCN and UCN with respect to the single-step neutron
conversion. A particularly promising medium is the weakly neutron absorbing,
fully deuterated type-II clathrate hydrate stabilized by molecular oxygen.
Its magnetic excitation at $0.4\ \mathrm{meV}$ is well placed to turn
neutrons from a cold source into VCN or UCN in a cascade of some dozen
collisions. A helpful peculiarity is the large Bragg cutoff wavelength of
the clathrate crystallites, $\lambda _{\mathrm{B}}\approx 2.0\ \mathrm{nm}$;
neutrons that still can impart kinetic energy to the moderator, i.e.\ those
with $E>E^{\ast }$ and correspondingly $\lambda <1.4\ \mathrm{nm}$, will be
confined in the moderator by strong diffusion. Also very helpful from a
practical point of view is the fact that neutron up-scattering becomes
insignificant already at ordinary liquid helium temperatures.

For a neutron spectrum prepared by a common cold source and for an inelastic
diffusion length in the order of $10\ \mathrm{cm}$, the paramagnetic cooling
cascade will take effect in a moderator with linear dimensions less than a
meter. This size should be considered as an upper limit since non-magnetic
degrees of freedom were neglected, leading to an underestimate of the true
moderation efficiency. Indeed, the O$_{2}$-hydrate possesses many
excitations on different energy scales, including rotations and librations
of encaged O$_{2}$ molecules, and the host lattice modes. The low-energy
excitations might offer shortcut channels for a faster, less space demanding
moderation of a precooled spectrum, thereby limiting the scope of the
paramagnetic cooling cascade to few scattering steps at low energy, where
the cross section is only slightly reduced by the magnetic form factor. A
candidate is the low-energy band of local, Einstein oscillator type modes
observed around $4.8\ \mathrm{meV}$ \cite{Chazallon2/2002}. If on the other
hand excitations at higher energies are sufficiently effective as well,
external premoderation to subthermal neutron temperatures might be
unnecessary. Otherwise there is also the viable option to couple the O$_{2}$%
-hydrate to a premoderator made of an advanced cold moderator medium, such
as solid methane, methane clathrate or mesitylene, all offering better
thermalization at low temperature than liquid hydrogen or deuterium \cite%
{Conrad/2004}.

With its aforementioned size, an O$_{2}$-hydrate moderator would fit in
thermal columns of TRIGA \cite{Fouquet/2003,Golub/1984}, PULSTAR \cite%
{Korobkina/2007} or WWR \cite{Serebrov/2009} type reactor facilities. Also
small accelerator based neutron facilities might offer excellent
opportunities for study and exploitation \cite{Lavelle/2008, Masuda/2002}.
Preparatory studies on neutron conversion can be performed at a neutron beam
similar to the experiments described in refs.\ \cite%
{Atchison/2005,Ageron/1978}. By exposing a larger quantity of the
paramagnetic medium to the beam one may also demonstrate the cascade gain
factor. Cross section measurements for different cage filling fractions
(which can be increased beyond $90$\% by a larger pressure of O$_{2}$ during
preparation of the clathrate \cite{Chazallon/2002}) can tell us if filling
with more than one O$_{2}$ molecule is a problem or an opportunity for
further increasing the moderation efficiency. A further point to be studied
is the transparency of the medium for neutrons having reached wavelengths $%
\lambda >\lambda _{\mathrm{B}}$ after the final cooling step. This will
depend on the level of inhomogeneity scattering \cite{Steyerl/1974} due to
the mesostructure, determined by size, packing and microporosity of
clathrate grains \cite{Kuhs/2000}. For implementation of the material in
intense neutron fields also the question of radiation damage needs to be
addressed.

An "in-pile" O$_{2}$-hydrate moderator may provide highest fluxes of VCN and
UCN. Highest ultimate UCN densities might however be attained in pure
superfluid $^{4}$He due to its vanishing neutron absorption, enabling UCN
accumulation prior to extraction from the converter \cite{Zimmer/2007}. Also
in this situation the clathrate may become an asset, as a moderating
reflector around a superfluid $^{4}$He UCN source placed at the end of a
neutron guide \cite{Baessler/2011,Bangert/1998}. Keeping the clathrate at $%
6\ \mathrm{K}$, the spectrum of an incident neutron beam will be compressed
to provide an optimum density at $1\ \mathrm{meV}$, where the one-phonon
process for UCN production in superfluid $^{4}$He takes effect.

Among the other media considered in this paper, also the dry O$_{2}$-$^{4}$%
He van der Waals cluster system deserves further investigation for its much
lower absorption. As this material becomes unstable at much lower
temperatures than the O$_{2}$-hydrate (see a phase diagram for the similar N$%
_{2}$-hydrate in ref.\ \cite{Kuhs/1997}), it might be rather a candidate for
implementation at a neutron beam than in-pile. Even more exotic are the
paramagnetic atomic Zeeman systems, which however offer a linear dependence
of the neutron transfer energy on an applied magnetic field. This degree of
freedom might be exploitable in some special experimental situations.

The absence of dispersion in neutron scattering by paramagnetic systems has
the additional interesting consequence that neutron conversion to UCN takes
place in a narrow energy range of a fraction of a $\mathrm{\mu eV}$. This
strongly contrasts with the dispersive single-phonon emission process in
superfluid $^{4}$He, which, for instance for a spectrum up to a cutoff at $%
E_{\mathrm{c}}=250\ \mathrm{neV}$, is kinematically allowed in a wide energy
range of $26\ \mathrm{\mu eV}$ about $1\ \mathrm{meV}$ \cite{Yoshiki/2003}.
While for the helium case the large mean free path of neutrons with energy $%
1\ \mathrm{meV}$ can in principle be used to enhance the UCN density by a
neutron beam resonator \cite{Zimmer/2014,Jericha/2000}, the paramagnetic
moderator with its narrow energy range for neutron conversion is amenable to
beam bunching techniques applicable for pulsed neutron sources, as described
in ref.\ \cite{Rauch/1994}. We finally note that it might also be worthwhile
to consider the combination of paramagnetic neutron cooling with Namiot's
original proposal, thus putting into practice a two-stage neutron cascade
cooler in which the spin dependent nuclear scattering compresses the neutron
phase space at lowest energies.

\section{Appendix: Neutron scattering cross sections}

In this appendix we derive the inelastic neutron scattering cross sections
needed for the analysis of neutron conversion and cascade cooling by
paramagnetic centers. The first part covers simple Zeeman systems of atomic
or ionic paramagnetic centers without zero-field splittings. The second part
deals with the triplet state of molecular oxygen without external magnetic
field. The analysis follows standard procedures presented in textbooks on
neutron scattering theory \cite{Lovesey/1984,Squires/1978} up to the point,
where we evaluate the thermal averages of time-dependent spin operators
without neglecting energy transfers to or from the neutron. While this can
in fact be easily accomplished for paramagnetic systems, expressions for
such inelastic cross sections seem not to appear in the literature, probably
because the usually small energy transfer in the diffuse scattering
associated with an electron spin flip is only of limited interest for
structural studies. As argued in the main text, the inelastic neutron
scattering due to the zero-field splitting in oxygen seems to have already
been observed in two experimental studies \cite{Chazallon2/2002,Renker/2001}%
, where it was however temptatively interpreted as a crystal field effect.
Also for this reason a comprehensive presentation of the corresponding cross
sections seems useful.

\subsection{Spin dependent neutron scattering cross sections for a Zeeman
system without zero-field splittings}

We analyze neutron scattering by atomic or ionic paramagnetic centers
polarized in a static external magnetic field and derive partial cross
sections for electron spin flip and electron non-spin flip processes, with
and without neutron spin flip. We start from the double differential cross
section for magnetic neutron scattering in first order time dependent
perturbation theory, which is given by%
\begin{equation}
\left( \frac{d^{2}\sigma }{d\Omega dE^{\prime }}\right) _{\eta \rightarrow
\eta ^{\prime }}=\frac{k^{\prime }}{k}\left( \frac{m_{\mathrm{n}}}{2\pi
\hbar ^{2}}\right) ^{2}\sum_{\lambda \lambda ^{\prime }}p_{\lambda
}\left\vert \left\langle \mathbf{k}^{\prime }\eta ^{\prime }\lambda ^{\prime
}\right\vert \mathcal{H}_{\mathrm{m}}\left( \mathbf{r}\right) \left\vert 
\mathbf{k}\eta \lambda \right\rangle \right\vert ^{2}\delta \left(
E_{\lambda ^{\prime }}-E_{\lambda }+E^{\prime }-E\right) .  \label{sigma}
\end{equation}%
Here a neutron with mass $m_{\mathrm{n}}$, wavevector $\mathbf{k}$, kinetic
energy $E$ and quantum number $\eta $ for the projection of the neutron spin
onto the $z$ axis defined by the external, static magnetic field $\mathbf{B}%
_{0}=\left( 0,0,B_{0}\right) $, is scattered into a final state with $%
\mathbf{k}^{\prime }$, $E^{\prime }$ and $\eta ^{\prime }$. The probed
system undergoes a transition from an initial state $\left\vert \lambda
\right\rangle $ characterized by a set of quantum numbers $\lambda $ and
energy $E_{\lambda }$ to a final state characterized by $\lambda ^{\prime }$
and energy $E_{\lambda ^{\prime }}$. The cross section in eq.\ \ref{sigma}
includes a sum over final states $\lambda ^{\prime }$ and thermal averaging
over the initial states by means of statistical weight factors $p_{\lambda }$%
.

The Hamiltonian $\mathcal{H}_{\mathrm{m}}\left( \mathbf{r}\right) =-\mathbf{%
\mu }_{\mathrm{n}}\cdot \mathbf{B}\left( \mathbf{r}\right) $ of the
interaction of the neutron magnetic moment $\mathbf{\mu }_{\mathrm{n}}=g_{%
\mathrm{n}}\mu _{\mathrm{N}}\mathbf{\sigma }/2$ with the local magnetic
field $\mathbf{B}\left( \mathbf{r}\right) $ in the paramagnetic system has
matrix elements between plane wave states $\mathbf{k}$ and $\mathbf{k}%
^{\prime }$ that can be expressed as%
\begin{equation}
\mathcal{H}_{\mathrm{m}}\left( \mathbf{\kappa }\right) =\left\langle \mathbf{%
k}^{\prime }\left\vert \mathcal{H}_{\mathrm{m}}\left( \mathbf{r}\right)
\right\vert \mathbf{k}\right\rangle =\frac{1}{2}\mu _{0}g_{\mathrm{n}}\mu _{%
\mathrm{N}}g_{\mathrm{e}}\mu _{\mathrm{B}}\mathbf{\sigma }\cdot \mathbf{Q}%
_{\bot }\left( \mathbf{\kappa }\right) ,  \label{H_m}
\end{equation}%
where $\mu _{0}$ is the magnetic vacuum permeability, $g_{\mathrm{n}}\approx
-3.826$ is the g-factor of the neutron, $\mu _{\mathrm{N}}$ is the nuclear
magneton, $g_{\mathrm{e}}\approx -2.002$ is the g-factor of the electron, $%
\mu _{\mathrm{B}}$ is the Bohr magneton, $\mathbf{\sigma }/2$ is the neutron
spin in units of $\hbar $ expressed by the vector of Pauli matrices $\mathbf{%
\sigma }=\left( \sigma _{x},\sigma _{y},\sigma _{z}\right) $, and the
scattering vector%
\begin{equation}
\mathbf{\kappa =k-k}^{\prime }
\end{equation}%
is the momentum transfer to the scattering system in units of $\hbar $. The
vector%
\begin{equation}
\mathbf{Q}_{\bot }=\widehat{\mathbf{\kappa }}\mathbf{\times }\left( \mathbf{%
Q\times }\widehat{\mathbf{\kappa }}\right) =\mathbf{Q-}\left( \mathbf{Q\cdot 
}\widehat{\mathbf{\kappa }}\right) \widehat{\mathbf{\kappa }}  \label{Q-perp}
\end{equation}%
is the component of a vector $\mathbf{Q}$ perpendicular to $\mathbf{\kappa }$
($\widehat{\mathbf{\kappa }}$ is the unit vector of $\mathbf{\kappa }$),
which can be shown to be in general proportional to the Fourier transform of
the atomic magnetization $\mathbf{M}\left( \mathbf{r}\right) $ due to both,
spin and orbital angular momentum of the unpaired electrons. We can limit
our attention to the case where unpaired electrons are located close to
equilibrium positions of paramagnetic centers, and where individual electron
spins of the center $j$ couple to a total spin $\mathbf{S}_{j}$ with quantum
number $S$. For the weakly absorbing species quoted in Table \ref{Species}
the total orbital angular momentum $\mathbf{L}_{j}$ vanishes. For low-energy
neutron scattering $S$ is a conserved quantum number while its $z$
component, characterized by a quantum number $m$, may change by one unit.
Under these circumstances the vector $\mathbf{Q}$ can be shown to take the
form%
\begin{equation}
\mathbf{Q=}\sum_{j}\mathbf{Q}_{j}\ \mathbf{=}\sum_{j}F_{j}\left( \mathbf{%
\kappa }\right) \exp \left( i\mathbf{\kappa \cdot R}_{j}\right) \mathbf{S}%
_{j},  \label{Q}
\end{equation}%
wherein $\mathbf{R}_{j}$ denotes the position of the $j$th paramagnetic
center and%
\begin{equation}
F_{j}\left( \mathbf{\kappa }\right) =\int \widetilde{s}_{j}\left( \mathbf{r}%
\right) \exp \left( i\mathbf{\kappa \cdot r}\right) d^{3}r
\end{equation}%
is the magnetic form factor with $\widetilde{s}_{j}$ denoting the density of
unpaired electrons of the $j$th ion, divided by their number, so that $%
F_{j}\left( 0\right) =1$.

The cross section is given in eq.\ \ref{sigma} for specific transitions
between neutron spin states $\left\vert +\right\rangle $ and $\left\vert
-\right\rangle $ with respect to the external magnetic field. From the
standard representation of the Pauli matrices the corresponding matrix
elements follow as 
\begin{equation}
\left\langle +\right\vert \mathbf{\sigma }\cdot \mathbf{Q}_{\bot }\left\vert
+\right\rangle =Q_{\bot z},\qquad \left\langle -\right\vert \mathbf{\sigma }%
\cdot \mathbf{Q}_{\bot }\left\vert -\right\rangle =-Q_{\bot z},  \label{sQ1}
\end{equation}%
and%
\begin{equation}
\left\langle -\right\vert \mathbf{\sigma }\cdot \mathbf{Q}_{\bot }\left\vert
+\right\rangle =Q_{\bot x}+iQ_{\bot y},\qquad \left\langle +\right\vert 
\mathbf{\sigma }\cdot \mathbf{Q}_{\bot }\left\vert -\right\rangle =Q_{\bot
x}-iQ_{\bot y},  \label{sQ2}
\end{equation}%
with the first (second) pair describing transitions without (with) neutron
spin flip. Considering first the cross sections for magnetic neutron spin
flip scattering, we use eqs.\ \ref{H_m} and \ref{sQ2} in eq.\ \ref{sigma}
and write%
\begin{equation}
\left( \frac{d^{2}\sigma }{d\Omega dE^{\prime }}\right) _{\pm \rightarrow
\mp }=b_{\mathrm{m}}^{2}\frac{k^{\prime }}{k}\sum_{\lambda \lambda ^{\prime
}}p_{\lambda }\left\langle \lambda \right\vert Q_{\bot x}^{\dag }\mp
iQ_{\bot y}^{\dag }\left\vert \lambda ^{\prime }\right\rangle \left\langle
\lambda ^{\prime }\right\vert Q_{\bot x}\pm iQ_{\bot y}\left\vert \lambda
\right\rangle \delta \left( E_{\lambda ^{\prime }}-E_{\lambda }+E^{\prime
}-E\right) ,  \label{sigma-Q}
\end{equation}%
where%
\begin{equation}
b_{\mathrm{m}}=\frac{1}{2}\mu _{0}g_{\mathrm{n}}\mu _{\mathrm{N}}g_{\mathrm{e%
}}\mu _{\mathrm{B}}\frac{m_{\mathrm{n}}}{2\pi \hbar ^{2}}=5.404\ \mathrm{fm}
\label{b_m}
\end{equation}%
is the magnetic scattering length. Continuing to follow the standard
procedure to evaluate the cross section, the $\delta $ function is expressed
as%
\begin{equation}
\delta \left( E_{\lambda ^{\prime }}-E_{\lambda }+E^{\prime }-E\right) =%
\frac{1}{2\pi \hbar }\int_{-\infty }^{\infty }\exp \left( i\left( E_{\lambda
^{\prime }}-E_{\lambda }\right) t/\hbar \right) \exp \left( i\left(
E^{\prime }-E\right) t/\hbar \right) dt.  \label{delta-int}
\end{equation}%
Since $\left\vert \lambda \right\rangle $ are eigenstates of the
Hamiltionian $\mathcal{H}_{0}$ of the system,%
\begin{equation}
\exp \left( i\mathcal{H}_{0}t/\hbar \right) \left\vert \lambda \right\rangle
=\exp \left( iE_{\lambda }t/\hbar \right) \left\vert \lambda \right\rangle .
\end{equation}%
One can define time dependent operators as%
\begin{equation}
Q_{\bot \alpha }\left( t\right) =\exp \left( i\mathcal{H}_{0}t/\hbar \right)
Q_{\bot \alpha }\exp \left( -i\mathcal{H}_{0}t/\hbar \right) ,
\end{equation}%
where $\alpha =x,y,z$ are cartesian coordinates with the $z$ axis pointing
along the external magnetic field. Using eq.\ \ref{Q} with this definition,
one can write%
\begin{equation}
Q_{\bot \alpha }\left( t\right) =\sum_{j}F_{j}\left( \mathbf{\kappa }\right)
\exp \left( i\mathbf{\kappa \cdot R}_{j}\left( t\right) \right) S_{\bot
j\alpha }\left( t\right) ,  \label{Q_perp-alpha}
\end{equation}%
where%
\begin{equation}
\mathbf{S}_{\bot j}\left( t\right) =\mathbf{S}_{j}\left( t\right) \mathbf{-}%
\left( \mathbf{S}_{j}\left( t\right) \mathbf{\cdot }\widehat{\mathbf{\kappa }%
}\right) \widehat{\mathbf{\kappa }},  \label{S-perp}
\end{equation}%
in analogy to eq.\ \ref{Q-perp}. Under the usual assumption that the
orientations of the electron spins do not affect positions and motion of the
nuclei, the thermal averages can be factorized for the nuclear coordinates
and electron spins. Using also the closure relation $\sum \left\vert \lambda
^{\prime }\right\rangle \left\langle \lambda ^{\prime }\right\vert =1$ and
denoting the thermal average $\sum p_{\lambda }\left\langle \lambda
\left\vert ...\right\vert \lambda \right\rangle $ by brackets $\left\langle
...\right\rangle $, the cross section becomes%
\begin{eqnarray}
\left( \frac{d^{2}\sigma }{d\Omega dE^{\prime }}\right) _{\pm \rightarrow
\mp } &=&\frac{b_{\mathrm{m}}^{2}}{2\pi \hbar }\frac{k^{\prime }}{k}%
\int_{-\infty }^{\infty }\sum_{jj^{\prime }}\left\langle \exp \left( -i%
\mathbf{\kappa \cdot R}_{j^{\prime }}\right) \exp \left( i\mathbf{\kappa
\cdot R}_{j}\left( t\right) \right) \right\rangle F_{j^{\prime }}^{\ast
}\left( \mathbf{\kappa }\right) F_{j}\left( \mathbf{\kappa }\right) 
\nonumber \\
&&\times \left\langle \left( S_{\bot j^{\prime }x}\mp iS_{\bot j^{\prime
}y}\right) \left( S_{\bot jx}\left( t\right) \pm iS_{\bot jy}\left( t\right)
\right) \right\rangle \exp \left( i\left( E^{\prime }-E\right) t/\hbar
\right) dt.
\end{eqnarray}%
It will be useful to employ the raising and lowering operators defined by%
\begin{equation}
S_{j}^{\pm }=S_{jx}\pm iS_{jy},  \label{S+-}
\end{equation}%
which fulfill the relation%
\begin{equation}
S_{j^{\prime }x}S_{jx}\left( t\right) +S_{j^{\prime }y}S_{jy}\left( t\right)
=\frac{1}{2}\left( S_{j^{\prime }}^{+}S_{j}^{-}\left( t\right) +S_{j^{\prime
}}^{-}S_{j}^{+}\left( t\right) \right) .  \label{xx-yy}
\end{equation}%
For further evaluation of the spin operator products in the cross section
one notes that for a paramagnetic system in an external magnetic field
applied in $z$\ direction, the total $z$\ component of the electron spin is
a constant of motion, and therefore\emph{\ }%
\begin{equation}
\sum_{j}\left[ S_{jz},\mathcal{H}_{0}\right] =0.  \label{com-Sz,H}
\end{equation}%
The operators $S_{j}^{\pm }$ then change the $z$ component of the total spin
of the system by one unit so that 
\begin{equation}
\left\langle S_{j^{\prime }}^{+}S_{j}^{+}\left( t\right) \right\rangle
=\left\langle S_{j^{\prime }}^{-}S_{j}^{-}\left( t\right) \right\rangle =0,
\end{equation}%
and therefore also%
\begin{equation}
\left\langle S_{j^{\prime }x}S_{jy}\left( t\right) +S_{j^{\prime
}y}S_{jx}\left( t\right) \right\rangle =\frac{1}{2i}\left\langle
S_{j^{\prime }}^{+}S_{j}^{+}\left( t\right) +S_{j^{\prime
}}^{-}S_{j}^{-}\left( t\right) \right\rangle =0.  \label{S-xy}
\end{equation}%
Also,%
\begin{equation}
\left\langle S_{j^{\prime }z}S_{jx}\left( t\right) \right\rangle =0,\qquad
\left\langle S_{j^{\prime }z}S_{jy}\left( t\right) \right\rangle =0,
\label{S-az}
\end{equation}%
and due to equivalence of the $x$ and $y$ axes,%
\begin{equation}
\left\langle S_{j^{\prime }x}S_{jx}\left( t\right) \right\rangle
=\left\langle S_{j^{\prime }y}S_{jy}\left( t\right) \right\rangle .
\label{cylind}
\end{equation}%
Since for a paramagnet there are no correlations between spins of different
centers $j\neq j^{\prime }$,%
\begin{equation}
\left\langle S_{j^{\prime }\alpha }S_{j\alpha }\left( t\right) \right\rangle
=\left\langle S_{j^{\prime }\alpha }\right\rangle \left\langle S_{j\alpha
}\right\rangle +\delta _{jj^{\prime }}\left( \left\langle S_{j\alpha
}S_{j\alpha }\left( t\right) \right\rangle -\left\langle S_{j^{\prime
}\alpha }\right\rangle \left\langle S_{j\alpha }\right\rangle \right) .
\end{equation}%
In presence of a static magnetic field in $z$ direction, $\left\langle
S_{jz}\right\rangle \neq 0$ but $\left\langle S_{jx}\right\rangle
=\left\langle S_{jy}\right\rangle =0$. The spin correlation functions
entering the cross section are thus given by%
\begin{equation}
\left\langle S_{j^{\prime }x}S_{jx}\left( t\right) \right\rangle
=\left\langle S_{j^{\prime }y}S_{jy}\left( t\right) \right\rangle =\delta
_{jj^{\prime }}\left\langle S_{x}S_{x}\left( t\right) \right\rangle ,
\end{equation}%
and%
\begin{equation}
\left\langle S_{j^{\prime }z}S_{jz}\left( t\right) \right\rangle
=\left\langle S_{z}\right\rangle ^{2}+\delta _{jj^{\prime }}\left(
\left\langle S_{z}S_{z}\left( t\right) \right\rangle -\left\langle
S_{z}\right\rangle ^{2}\right) ,  \label{S-corr}
\end{equation}%
where by omission of the index $j$ we focus attention on a medium containing
a single paramagnetic species without anisotropy effects due to
electrostatic crystal fields. The cross section for neutron spin flip
scattering thus becomes 
\begin{eqnarray}
\left( \frac{d^{2}\sigma }{d\Omega dE^{\prime }}\right) _{\eta \neq \eta
^{\prime }} &=&\frac{b_{\mathrm{m}}^{2}}{2\pi \hbar }\frac{k^{\prime }}{k}%
\int_{-\infty }^{\infty }\sum_{jj^{\prime }}\left\langle \exp \left( -i%
\mathbf{\kappa \cdot R}_{j^{\prime }}\right) \exp \left( i\mathbf{\kappa
\cdot R}_{j}\left( t\right) \right) \right\rangle \left\vert F\left( \mathbf{%
\kappa }\right) \right\vert ^{2}  \nonumber \\
&&\times \left[ \delta _{jj^{\prime }}\left( \frac{1}{4}\left( 1+\widehat{%
\kappa }_{z}^{4}\right) \left\langle S^{+}S^{-}\left( t\right)
+S^{-}S^{+}\left( t\right) \right\rangle +\left( \widehat{\kappa }_{z}^{2}-%
\widehat{\kappa }_{z}^{4}\right) \left( \left\langle S_{z}S_{z}\left(
t\right) \right\rangle -\left\langle S_{z}\right\rangle ^{2}\right) \right)
\right.  \nonumber \\
&&\left. +\left( \widehat{\kappa }_{z}^{2}-\widehat{\kappa }_{z}^{4}\right)
\left\langle S_{z}\right\rangle ^{2}\right] \times \exp \left( i\left(
E^{\prime }-E\right) t/\hbar \right) dt.  \label{sigma-S-para}
\end{eqnarray}%
where we have written $\eta \neq \eta ^{\prime }$ instead of $\pm
\rightarrow \mp $, since the cross section is found to be independent on the
neutron's spin flipping from up to down or vice versa, in contrast to
nuclear scattering by polarized nuclei. The cross section for magnetic
neutron non spin flip scattering can be derived accordingly, with the
replacement of the matrix element product in eq.\ \ref{sigma-Q} by $%
\left\langle \lambda \right\vert Q_{\bot z}^{\dagger }\left\vert \lambda
^{\prime }\right\rangle \left\langle \lambda ^{\prime }\right\vert Q_{\bot
z}\left\vert \lambda \right\rangle $. This results in%
\begin{eqnarray}
\left( \frac{d^{2}\sigma }{d\Omega dE^{\prime }}\right) _{\eta =\eta
^{\prime }} &=&\frac{b_{\mathrm{m}}^{2}}{2\pi \hbar }\frac{k^{\prime }}{k}%
\int_{-\infty }^{\infty }\sum_{jj^{\prime }}\left\langle \exp \left( -i%
\mathbf{\kappa \cdot R}_{j^{\prime }}\right) \exp \left( i\mathbf{\kappa
\cdot R}_{j}\left( t\right) \right) \right\rangle \left\vert F\left( \mathbf{%
\kappa }\right) \right\vert ^{2}  \nonumber \\
&&\times \left[ \delta _{jj^{\prime }}\left( \frac{1}{4}\left( \widehat{%
\kappa }_{z}^{2}-\widehat{\kappa }_{z}^{4}\right) \left\langle
S^{+}S^{-}\left( t\right) +S^{-}S^{+}\left( t\right) \right\rangle +\left( 1-%
\widehat{\kappa }_{z}^{2}\right) ^{2}\left( \left\langle S_{z}S_{z}\left(
t\right) \right\rangle -\left\langle S_{z}\right\rangle ^{2}\right) \right)
\right.  \nonumber \\
&&\left. +\left( 1-\widehat{\kappa }_{z}^{2}\right) ^{2}\left\langle
S_{z}\right\rangle ^{2}\right] \times \exp \left( i\left( E^{\prime
}-E\right) t/\hbar \right) dt.  \label{sigma-S'-para}
\end{eqnarray}

The time dependence of the spin observables is governed by the Hamiltonian
of a paramagnetic center in the external magnetic field, i.e.%
\begin{equation}
\mathcal{H}_{0}=-g\mu _{\mathrm{B}}B_{0}S_{z}.  \label{H_0}
\end{equation}%
The energy levels are given by the eigenstates of $S_{z}$ with quantum
number $m$, 
\begin{equation}
\mathcal{H}_{0}\left\vert m\right\rangle =E_{m}\left\vert m\right\rangle ,
\label{H_0m}
\end{equation}%
with%
\begin{equation}
E_{m}=-g\mu _{\mathrm{B}}B_{0}m.  \label{E_m}
\end{equation}%
The g-factors of the paramagnetic centers listed in Table \ref{Species} are $%
g\approx -2$. Energy transfers to or from the neutron may occur in units of
the Zeeman energy denoted as%
\begin{equation}
E^{\ast }=\left\vert g\mu _{\mathrm{B}}B_{0}\right\vert .
\end{equation}%
The system in thermal equilibrium at temperature $T$ is chacterized by a
partition function $Z$, with the population probabilities of the states $%
\left\vert m\right\rangle $ given by%
\begin{equation}
p_{m}=\frac{\exp \left( -\beta E_{m}\right) }{Z},\qquad Z=\sum_{m}\exp
\left( -\beta E_{m}\right) ,  \label{p_m and Z}
\end{equation}%
where the sum extends over the values $-S\leq m\leq S$, and%
\begin{equation}
\beta =\left( k_{\mathrm{B}}T\right) ^{-1}  \label{beta}
\end{equation}%
with the Boltzmann constant $k_{\mathrm{B}}$.

Evaluating first the matrix elements of operators $S_{z}$ in eqs.\ \ref%
{sigma-S-para} and \ref{sigma-S'-para}, we note that%
\begin{equation}
\left\langle S_{z}\right\rangle =\sum_{m}p_{m}\left\langle m\right\vert
S_{z}\left\vert m\right\rangle =\sum_{m}p_{m}m  \label{Sz}
\end{equation}%
and%
\begin{equation}
\left\langle S_{z}S_{z}\left( t\right) \right\rangle
=\sum_{m}p_{m}\left\langle m\right\vert S_{z}S_{z}\left( t\right) \left\vert
m\right\rangle =\sum_{m}p_{m}m^{2}=\left\langle S_{z}^{2}\right\rangle
\label{S2z}
\end{equation}%
are both time independent and thus describe scattering without electronic
spin flip. The partition function of the Zeeman system is given by%
\begin{equation}
Z=\sum_{m}\exp \left( -mx\right) =\frac{\sinh \left( \left( S+\frac{1}{2}%
\right) x\right) }{\sinh \frac{x}{2}},\qquad x=-\beta g\mu _{\mathrm{B}%
}B_{0},  \label{Z}
\end{equation}%
from which, using eqs.\ \ref{Sz} and \ref{S2z}, follow the thermal average
values of the spin observables $S_{z}$ and $S_{z}^{2}$ as%
\begin{eqnarray}
\left\langle S_{z}\right\rangle &=&\frac{1}{Z}\sum_{m}m\exp \left(
-mx\right) =-\frac{1}{Z}\frac{dZ}{dx}  \label{Sz-mean} \\
&=&-\frac{1}{2}\left( \left( 2S+1\right) \coth \left( \frac{x}{2}\left(
2S+1\right) \right) -\coth \frac{x}{2}\right) ,  \nonumber
\end{eqnarray}%
and%
\begin{eqnarray}
\left\langle S_{z}^{2}\right\rangle &=&\frac{1}{Z}\sum_{m}m^{2}\exp \left(
-mx\right) =\frac{1}{Z}\frac{d^{2}Z}{dx^{2}}  \label{Sz2-mean} \\
&=&S\left( S+1\right) +\left\langle S_{z}\right\rangle \coth \frac{x}{2}. 
\nonumber
\end{eqnarray}

Next we analyze the matrix elements involving the operators $S^{\pm }$ in
eqs.\ \ref{sigma-S-para} and \ref{sigma-S'-para}. Application of the time
independent operators to a state with quantum number $m$ results in%
\begin{equation}
S^{\pm }\left\vert m\right\rangle =\sqrt{S\left( S+1\right) -m\left( m\pm
1\right) }\left\vert m\pm 1\right\rangle .  \label{S+-m}
\end{equation}%
Employing the time dependent operators%
\begin{equation}
S^{\pm }\left( t\right) =\exp \left( i\mathcal{H}_{0}t/\hbar \right) S^{\pm
}\exp \left( -i\mathcal{H}_{0}t/\hbar \right) ,  \label{S+-(t)}
\end{equation}%
and using eq.\ \ref{S+-m} and eq.\ \ref{S+-(t)} with eq.\ \ref{H_0m}, we
obtain 
\begin{eqnarray}
\left\langle S^{\pm }S^{\mp }\left( t\right) \right\rangle
&=&\sum_{m}p_{m}\left\langle m\right\vert S^{\pm }S^{\mp }\left( t\right)
\left\vert m\right\rangle  \label{S+-S-+} \\
&=&\sum_{m}p_{m}\left( S\left( S+1\right) -m\left( m\mp 1\right) \right)
\exp \left( i\left( E_{m\mp 1}-E_{m}\right) t/\hbar \right) .  \nonumber
\end{eqnarray}%
These thermal averages thus describe electronic spin flips and associated
energy transfer from or to the neutron. Using eqs.\ \ref{Sz} and \ref{S2z}
they can be expressed as%
\begin{equation}
\left\langle S^{\pm }S^{\mp }\left( t\right) \right\rangle =\left( S\left(
S+1\right) -\left\langle S_{z}^{2}\right\rangle \pm \left\langle
S_{z}\right\rangle \right) \exp \left( \pm ig\mu _{\mathrm{B}}B_{0}t/\hbar
\right) .  \label{S+S--mean}
\end{equation}%
with the explicit temperature dependences of the thermal averages given in
eqs. \ref{Sz-mean} and \ref{Sz2-mean}.

The cross sections given in eqs.\ \ref{sigma-S-para} and \ref{sigma-S'-para}
can now be evaluated, using eqs.\ \ref{xx-yy},\ \ref{S2z} and \ref{S+S--mean}%
, with integration over time and collecting the terms that correspond to
electronic spin flip and those which do not. We denote the partial cross
sections with electronic spin flip leading to a loss (gain) in neutron
energy by a superscript $-$ $\left( +\right) $, and those without electronic
spin flip by a superscript $0$, i.e. 
\begin{equation}
\left( \frac{d^{2}\sigma }{d\Omega dE^{\prime }}\right) _{\eta \neq \eta
^{\prime }}^{\pm }=\frac{A}{4}\left( 1+\widehat{\kappa }_{z}^{4}\right)
\left( S\left( S+1\right) -\left\langle S_{z}^{2}\right\rangle \pm
\left\langle S_{z}\right\rangle \right) \delta \left( E-E^{\prime }\pm
E^{\ast }\right) ,  \label{sigma-1}
\end{equation}%
\begin{equation}
\left( \frac{d^{2}\sigma }{d\Omega dE^{\prime }}\right) _{\eta =\eta
^{\prime }}^{\pm }=\frac{A}{4}\left( \widehat{\kappa }_{z}^{2}-\widehat{%
\kappa }_{z}^{4}\right) \left( S\left( S+1\right) -\left\langle
S_{z}^{2}\right\rangle \pm \left\langle S_{z}\right\rangle \right) \delta
\left( E-E^{\prime }\pm E^{\ast }\right) ,  \label{sigma-2}
\end{equation}%
\begin{equation}
\left( \frac{d^{2}\sigma }{d\Omega dE^{\prime }}\right) _{\eta \neq \eta
^{\prime }}^{0}=A\left( \widehat{\kappa }_{z}^{2}-\widehat{\kappa }%
_{z}^{4}\right) \left( \left\langle S_{z}^{2}\right\rangle -\left\langle
S_{z}\right\rangle ^{2}+\left\langle S_{z}\right\rangle ^{2}\sum_{j}\exp
\left( i\mathbf{\kappa \cdot R}_{j}\right) \right) \delta \left( E^{\prime
}-E\right) ,  \label{sigma-3}
\end{equation}%
\begin{equation}
\left( \frac{d^{2}\sigma }{d\Omega dE^{\prime }}\right) _{\eta =\eta
^{\prime }}^{0}=A\left( 1-\widehat{\kappa }_{z}^{2}\right) ^{2}\left(
\left\langle S_{z}^{2}\right\rangle -\left\langle S_{z}\right\rangle
^{2}+\left\langle S_{z}\right\rangle ^{2}\sum_{j}\exp \left( i\mathbf{\kappa
\cdot R}_{j}\right) \right) \delta \left( E^{\prime }-E\right) .
\label{sigma-4}
\end{equation}%
The common factor%
\begin{equation}
A=Nb_{\mathrm{m}}^{2}\frac{k^{\prime }}{k}\exp \left( -2W\right) \left\vert
F\left( \mathbf{\kappa }\right) \right\vert ^{2},
\end{equation}%
contains the total number $N$ of paramagnetic centers, and the Debye-Waller
factor $\exp \left( -2W\right) $, where $2W=\kappa ^{2}\left\langle u_{%
\mathbf{\kappa }}^{2}\right\rangle $, and $\left\langle u_{\mathbf{\kappa }%
}^{2}\right\rangle $ is the mean square displacement of a paramagnetic
center in direction of $\mathbf{\kappa }$.

The cross sections in eqs.\ \ref{sigma-1} and \ref{sigma-2} involving an
electron spin flip with energy transfer $\pm E^{\ast }$ are incoherent; they
do not contain terms due to interferences of amplitudes from different
paramagnetic centers. They vanish if the energy of the incident neutron is
too small to compensate for the Zeeman energy needed to flip a single
electron spin. In the opposite limit, $E\gg E^{\ast }$, and if one is not
interested in the energy transfer, neglect of $E^{\ast }$ in the $\delta $
functions and summing up the partial cross sections for electron spin flip
and non-spin flip leads to equations found in the text books.

The electron non spin flip cross sections given in eqs.\ \ref{sigma-3} and %
\ref{sigma-4} describe elastic scattering (if neglecting the neutron Zeeman
energy in case of neutron spin flip scattering, the approximation adopted
here). They contain an incoherent diffuse term and a coherent term
proportional to $\left\langle S_{z}\right\rangle ^{2}$ due to interferences
of amplitudes from different paramagnetic centers, which may show up in
Bragg peaks, or lead to small angle scattering contrast for instance for
agglomerations of paramagnetic centers immersed in a non-magnetic solvent.
Another noteworthy feature is the fact that the coherent cross section with
neutron spin flip does not vanish in directions for which $\widehat{\kappa }%
_{z}^{2}-\widehat{\kappa }_{z}^{4}\neq 0$, i.e.\ when $\mathbf{\kappa }$
does not point parallel or perpendicular to the applied magnetic field.

For our calculations on neutron conversion and cooling we are primarily
interested in the neutron energy changing total cross sections. After
integration of $\widehat{\kappa }_{z}^{2}$ and $\widehat{\kappa }_{z}^{4}$
over solid angle,%
\begin{equation}
\int \widehat{\kappa }_{z}^{2}d\Omega =\frac{4}{3}\pi ,\qquad \int \widehat{%
\kappa }_{z}^{4}d\Omega =\frac{4}{5}\pi ,
\end{equation}%
we can write them as 
\begin{equation}
\left( \frac{d\sigma }{dE^{\prime }}\right) ^{\pm }=\left( \frac{d\sigma }{%
dE^{\prime }}\right) _{\eta \neq \eta ^{\prime }}^{\pm }+\left( \frac{%
d\sigma }{dE^{\prime }}\right) _{\eta =\eta ^{\prime }}^{\pm }=N\sigma _{%
\mathrm{m}}\frac{k^{\prime }}{k}\exp \left( -2W\right) g_{\pm }\left(
T\right) f_{\pm }\left( E\right) \delta \left( E\pm E^{\ast }-E^{\prime
}\right) ,  \label{dS/dE'}
\end{equation}%
where we have defined $\sigma _{\mathrm{m}}=4\pi b_{\mathrm{m}}^{2}\approx
3.66\ \mathrm{barn}$ and%
\begin{equation}
g_{\pm }\left( T\right) =\frac{1}{3}\left( S\left( S+1\right) -\left\langle
S_{z}^{2}\right\rangle \pm \left\langle S_{z}\right\rangle \right) ,
\end{equation}%
with $\left\langle S_{z}\right\rangle $ and $\left\langle
S_{z}^{2}\right\rangle $ given by eqs.\ \ref{Sz-mean} and \ref{Sz2-mean}.
The functions $f_{\pm }\left( E\right) $ account for the magnetic form
factor, which is discussed in the main text.

\subsection{Cross sections for the molecular oxygen spin triplet system}

Here we consider magnetic neutron scattering by an assembly of unoriented
oxygen molecules with motions frozen out. The molecules are assumed to be
kept sufficiently far apart from each other to avoid magnetic ordering. This
can be achieved using the cage structures discussed in the main text. Our
primary interest is the scattering involving transitions between magnetic
levels within the triplet state, which is inelastic due to the molecular
zero-field splitting. For unoriented molecules and without external magnetic
field there is no global quantization axis in the system. It is therefore
appropriate to start from the magnetic scattering cross section for
unpolarized neutrons (see, e.g.\ \cite{Squires/1978}),%
\begin{equation}
\frac{d^{2}\sigma }{d\Omega dE^{\prime }}=b_{\mathrm{m}}^{2}\frac{k^{\prime }%
}{k}\sum_{\alpha \beta }\left( \delta _{\alpha \beta }-\widehat{\kappa }%
_{\alpha }\widehat{\kappa }_{\beta }\right) \sum_{\lambda \lambda ^{\prime
}}p_{\lambda }\left\langle \lambda \right\vert Q_{\alpha }^{\dag }\left\vert
\lambda ^{\prime }\right\rangle \left\langle \lambda ^{\prime }\right\vert
Q_{\beta }\left\vert \lambda \right\rangle \delta \left( E_{\lambda ^{\prime
}}-E_{\lambda }+E^{\prime }-E\right) ,  \label{sigma-unpolarized}
\end{equation}%
using the same notation of states and transition operators as in the
previous section. Each oxygen molecule is characterized by a coordinate $%
\mathbf{R}_{j}$ of its center of gravity and relative positions $\mathbf{l}%
_{j1}$ and $\mathbf{l}_{j2}$ of the two atoms. Projection of the total spin
onto the molecular axis, $\mathbf{l}_{j1}-\mathbf{l}_{j2}$, provides a good
quantum number. As we do not deal with nuclear scattering, the atomic
coordinates do not explicitly occur as variables in the cross section but
manifest implicitly as a site dependence of the spin eigenstates. Also the
magnetic form factor depends on the molecular orientation, which we can
however take as isotropic for our purposes (see section $2$). Taking
electronic spins and spatial coordinates as independent quantities we write%
\begin{eqnarray}
\frac{d^{2}\sigma }{d\Omega dE^{\prime }} &=&\frac{b_{\mathrm{m}}^{2}}{2\pi
\hbar }\frac{k^{\prime }}{k}\left\vert F\left( \kappa \right) \right\vert
^{2}\sum_{\alpha \beta }\left( \delta _{\alpha \beta }-\widehat{\kappa }%
_{\alpha }\widehat{\kappa }_{\beta }\right) \times \\
&&\times \int_{-\infty }^{\infty }\sum_{jj^{\prime }}\left\langle \exp
\left( -i\mathbf{\kappa \cdot R}_{j^{\prime }}\right) \exp \left( i\mathbf{%
\kappa \cdot R}_{j}\left( t\right) \right) \right\rangle \left\langle
S_{j^{\prime }\alpha }S_{j\beta }\left( t\right) \right\rangle \exp \left(
i\left( E^{\prime }-E\right) t/\hbar \right) dt.  \nonumber
\end{eqnarray}%
For uncorrelated oxygen molecules,%
\begin{equation}
\left\langle S_{j^{\prime }\alpha }S_{j\beta }\left( t\right) \right\rangle
=\left\langle S_{j^{\prime }\alpha }\right\rangle \left\langle S_{j\beta
}\left( t\right) \right\rangle =0\qquad \left( j\neq j^{\prime }\right) .
\end{equation}%
We are thus left with a single sum over an assembly of unoriented and
independent triplet spin systems,%
\begin{eqnarray}
\frac{d^{2}\sigma }{d\Omega dE^{\prime }} &=&\frac{b_{\mathrm{m}}^{2}}{2\pi
\hbar }\frac{k^{\prime }}{k}\exp \left( -2W\right) \left\vert F\left( \kappa
\right) \right\vert ^{2}\times \\
&&\times \int_{-\infty }^{\infty }\sum_{j}\sum_{\alpha \beta }\left( \delta
_{\alpha \beta }-\widehat{\kappa }_{\alpha }\widehat{\kappa }_{\beta
}\right) \left\langle S_{j\alpha }S_{j\beta }\left( t\right) \right\rangle
\exp \left( i\left( E^{\prime }-E\right) t/\hbar \right) dt.  \nonumber
\end{eqnarray}%
The product $\left\langle \exp \left( -i\mathbf{\kappa \cdot R}_{j}\right)
\exp \left( i\mathbf{\kappa \cdot R}_{j}\left( t\right) \right)
\right\rangle $ is the Debye-Waller factor denoted by $\exp \left(
-2W\right) $. In the sum over $j$ any molecular orientation appears with
equal weight with respect to the given direction $\widehat{\mathbf{\kappa }}$%
, of which the differential cross section is obviously independent. We may
therefore define for each molecule its own coordinate system and replace%
\begin{equation}
\sum_{j}\sum_{\alpha \beta }\left( \delta _{\alpha \beta }-\widehat{\kappa }%
_{\alpha }\widehat{\kappa }_{\beta }\right) \left\langle S_{j\alpha
}S_{j\beta }\left( t\right) \right\rangle =N\left\langle \sum_{\alpha \beta
}\left( \delta _{\alpha \beta }-\widehat{\kappa }_{\alpha }\widehat{\kappa }%
_{\beta }\right) S_{\alpha }S_{\beta }\left( t\right) \right\rangle .
\label{ave}
\end{equation}%
On the right side the brackets include angular averaging in addition to the
thermal averaging over molecular spin states. Accordingly we have omitted
the site index $j$ to the spin operators. For further evaluation we choose
local cartesian coordinates with $z$ axis parallel to the molecular axis and
take the $x$ and $y$ axes in directions for which their projections on $%
\mathbf{\kappa }$ are equal, i.e.%
\begin{equation}
\widehat{\kappa }_{x}=\widehat{\kappa }_{y}=\frac{1}{\sqrt{2}}\sin \vartheta
,\qquad \widehat{\kappa }_{z}=\cos \vartheta ,
\end{equation}%
where $\vartheta $ is the angle between $\mathbf{\kappa }$ and the molecular
axis. The triplet states of the oxygen molecule are labelled by quantum
numbers $m=-1,0,+1$ characterizing the spin state projection along the
symmetry axis of the molecule. The Hamiltonian (without external magnetic
field) is given by%
\begin{equation}
\mathcal{H}_{0}=DS_{z}^{2}-\frac{2}{3}D,
\end{equation}%
which accounts for the energy difference by the zero-field splitting
constant $D$ of the states with $m=\pm 1$ and the $m=0$ state \cite%
{Wasserman/1964}. It commutes with $S_{z}$, and since the spin operators
obey the same algebra as in the Zeeman case (with different meaning of the
states), with the definition of raising and lowering operators in eq.\ \ref%
{S+-}, we use eqs.\ \ref{xx-yy}, \ref{S-xy}, \ref{S-az} and \ref{S2z}, and
obtain%
\begin{equation}
\left\langle \sum_{\alpha \beta }\left( \delta _{\alpha \beta }-\widehat{%
\kappa }_{\alpha }\widehat{\kappa }_{\beta }\right) S_{\alpha }S_{\beta
}\left( t\right) \right\rangle =\frac{1}{3}\left\langle \left(
S^{+}S^{-}\left( t\right) +S^{-}S^{+}\left( t\right) \right) \right\rangle
+\allowbreak \frac{2}{3}\left\langle S_{z}^{2}\right\rangle .
\end{equation}

The eigenenergies of $\mathcal{H}_{0}$ are given by%
\begin{equation}
E_{m}=\left( m^{2}-\frac{2}{3}\right) D.  \label{E_m  triplet}
\end{equation}%
The partition function as defined in eq.\ \ref{p_m and Z} follows as%
\begin{equation}
Z=\exp \left( 2x\right) +2\exp \left( -x\right) ,\qquad x=\frac{\beta D}{3}.
\end{equation}%
Using eq.\ \ref{S+-S-+} with eq.\ \ref{E_m triplet}, we obtain%
\begin{equation}
\left\langle S^{+}S^{-}\left( t\right) +S^{-}S^{+}\left( t\right)
\right\rangle =4\frac{\exp \left( iDt/\hbar \right) +\exp \left( -\beta
D\right) \exp \left( -iDt/\hbar \right) }{1+2\exp \left( -\beta D\right) }.
\end{equation}

The total scattering cross sections with neutron energy loss ($-$), energy
gain ($+$) thus become%
\begin{equation}
\left( \frac{d\sigma }{dE^{\prime }}\right) ^{\pm }=N\sigma _{\mathrm{m}}%
\frac{k^{\prime }}{k}\exp \left( -2W\right) g_{\pm }\left( T\right) f_{\pm
}\left( E\right) \delta \left( E\pm D-E^{\prime }\right) ,
\end{equation}%
where%
\begin{equation}
g_{-}\left( T\right) =\frac{4}{3}\frac{1}{1+2\exp \left( -\beta D\right) },
\end{equation}%
and%
\begin{equation}
g_{+}\left( T\right) =\frac{4}{3}\frac{\exp \left( -\beta D\right) }{1+2\exp
\left( -\beta D\right) }.
\end{equation}%
The functions $f_{\pm }\left( E\right) $ account for the magnetic form
factor as discussed in the main text. The cross sections fulfill the
relation of detailed balance, as they have to.

Using eq.\ \ref{S2z} we also obtain%
\begin{equation}
\left\langle S_{z}^{2}\right\rangle =\frac{2}{2+\exp \left( \beta D\right) },
\end{equation}%
from which follows the (for our present purposes less interesting) elastic
cross section as%
\begin{equation}
\left( \frac{d\sigma }{dE^{\prime }}\right) ^{0}=N\sigma _{\mathrm{m}}\exp
\left( -2W\right) g_{0}\left( T\right) f_{0}\left( E\right) \delta \left(
E-E^{\prime }\right) ,
\end{equation}%
where%
\begin{equation}
g_{0}\left( T\right) =\frac{4}{3}\frac{1}{2+\exp \left( \beta D\right) }%
\rightarrow 0\qquad \left( T\rightarrow 0\right) ,
\end{equation}%
and%
\begin{equation}
f_{0}\left( E\right) =\frac{1}{2}\int_{0}^{\pi }\left\langle \left\vert
F\right\vert ^{2}\right\rangle \left( \kappa _{0}\left( E,\theta \right)
\right) \sin \theta d\theta .
\end{equation}%
The brackets stand for orientational averaging of the molecules, and%
\begin{equation}
\kappa _{0}=\frac{2}{\hbar }\sqrt{m_{\mathrm{n}}E\left( 1-\cos \theta
\right) }.
\end{equation}

\end{document}